\documentclass[iop]{emulateapj}

\usepackage{color}
\usepackage{apjfonts}
\usepackage{natbib}
\usepackage[dvipdfmx,colorlinks=true,linkcolor=blue,filecolor=blue,urlcolor=blue,citecolor=blue]{hyperref}
\usepackage{amsmath}
\hypersetup
{
	pdfauthor={T. Izumi et al.},
	pdftitle={Izumi2015c}
}

\shorttitle{Do CNDs drive the accretion onto Seyfert nuclei?}
\shortauthors{T. Izumi et al.}

\begin{document}

\title{Do circumnuclear dense gas disks drive \\
mass accretion onto supermassive black holes?}


\author{Takuma Izumi\altaffilmark{1}, 
Nozomu Kawakatu\altaffilmark{2}, and 
Kotaro Kohno\altaffilmark{1,3}
}

\affil{${}^{1}$\ Institute of Astronomy, School of Science, The University of Tokyo, 2-21-1 Osawa, Mitaka, Tokyo 181-0015, Japan; \href{mailto:takumaizumi@ioa.s.u-tokyo.ac.jp}{takumaizumi@ioa.s.u-tokyo.ac.jp}}
\affil{${}^{2}$\ Faculty of Natural Sciences, National Institute of Technology, Kure College, 2-2-11 Agaminami, Kure, Hiroshima 737-8506, Japan} 
\affil{${}^{3}$\ Research Center for the Early Universe, The University of Tokyo, 7-3-1 Hongo, Bunkyo, Tokyo 113-0033}


\begin{abstract}
We present a positive correlation between the mass of dense molecular gas ($M_{\rm dense}$) 
of $\sim$ 100 pc scale circumnuclear disks (CNDs) 
and the black hole mass accretion rate ($\dot{M}_{\rm BH}$) in total 10 Seyfert galaxies, 
based on data compiled from the literature and an archive (median aperture $\theta_{\rm med}$ = 220 pc). 
A typical $M_{\rm dense}$ of CNDs is 10$^{7-8}$ $M_\odot$, 
estimated from the luminosity of the dense gas tracer, the HCN($1-0$) emission line. 
Because dense molecular gas is the site of star formation, 
this correlation is virtually equivalent to the one between 
nuclear star formation rate and $\dot{M}_{\rm BH}$ revealed previously. 
Moreover, the $M_{\rm dense}-\dot{M}_{\rm BH}$ correlation was tighter 
for CND-scale gas than for the gas on kpc or larger scales. 
This indicates that CNDs likely play an important role in fueling black holes, 
whereas $>$kpc scale gas does not. 
To demonstrate a possible approach for studying the CND-scale accretion process 
with the Atacama Large Millimeter/submillimeter Array (ALMA), 
we used a mass accretion model where angular momentum loss 
due to supernova explosions is vital. 
Based on the model prediction, we suggest that only the partial fraction of the mass 
accreted from the CND ($\dot{M}_{\rm acc}$) is consumed as $\dot{M}_{\rm BH}$. 
However, $\dot{M}_{\rm acc}$ agrees well with 
the total nuclear mass flow rate (i.e., $\dot{M}_{\rm BH}$ + outflow rate). 
Although these results are still tentative with large uncertainties, 
they support the view that star formation in CNDs 
can drive mass accretion onto supermassive black holes in Seyfert galaxies. 
\end{abstract}


\section{Introduction}\label{sec1} 
The mass accretion onto a supermassive black hole 
(SMBH, with a mass of $M_{\rm BH} \gtrsim 10^6$ $M_\odot$) 
is commonly believed to produce the enormous amount of energy 
observed as an active galactic nucleus (AGN; \citealt{1993ARA&A..31..473A,1995PASP..107..803U}). 
However, the physics of the angular momentum transfer 
of the accreting gas remain unclear (\citealt{2012NewAR..56...93A} for a review). 
SMBHs have been claimed to reside at the centers of galaxies with spheroidal components, 
and there is a correlation between $M_{\rm BH}$ and the properties of the host galaxies, 
such as bulge mass ($M_{\rm bulge}$) and stellar velocity dispersion ($\sigma_*$). 
This, a so called {\it{co-evolutionary}} relationship 
(\citealt{1998AJ....115.2285M,2000ApJ...539L...9F,2002ApJ...574..740T,2003ApJ...589L..21M,2009ApJ...698..198G}; 
\citealt{2013ARA&A..51..511K} and references therein), 
indicates that bulges and SMBHs evolved together, by regulating each other. 
Thus, to understand the mechanism of the angular momentum transfer of the accreting gas 
at various spatial scales (from a host galaxy to an accretion disk), 
which is directly connected to the mass accumulation of an SMBH, 
is of great importance to unveil the currently unknown co-evolutionary mechanism. 

Recent numerical simulations have shown that the radial streaming of gas caused by 
major mergers can efficiently feed the central SMBH, 
which triggers a powerful AGN as well as starbursts (e.g., \citealt{2008ApJS..175..356H,2010MNRAS.407.1529H}). 
While such a violent mechanism would be vital for fueling luminous quasars, 
rather secular processes induced by, for example, barred gravitational potential, 
galaxy$-$galaxy interactions, or even joint effects of star formation 
and large-scale dynamics (e.g., \citealt{1990Natur.345..679S,1992ARA&A..30..705B,
2004ARA&A..42..603K,2006ApJS..166....1H,2006LNP...693..143J,2015ApJ...803L...8R}), 
or minor mergers (e.g., \citealt{1994ApJ...425L..13M,1999ApJ...524...65T,2014MNRAS.440.2944K}), 
would be sufficient to explain lower-luminosity activity, such as that observed in Seyfert galaxies 
(e.g., \citealt{2009ApJ...694..599H,2012ApJ...758L..39T}). 
The lack of enhanced signatures of major mergers or strong interactions 
in local Seyfert galaxies supports this view (\citealt{2009ApJ...691..705G,2011ApJ...726...57C}). 

Although large-scale structures, like a bar, would efficiently 
transport gases towards the central region (e.g., \citealt{1999ApJ...525..691S,2005ApJ...632..217S}), 
there is little or no clear difference in that scale morphologies between AGN hosts and inactive galaxies 
(e.g., \citealt{1997ApJ...482L.135M,2003ApJ...589..774M,2007ApJ...655..718S,2015MNRAS.447..506C,2015ApJ...802..137C}). 
Even for smaller-scale morphologies, such as 
nuclear bars and nuclear spirals (e.g., \citealt{1990Natur.345..679S,2010MNRAS.407.1529H}), 
this trend holds, at least in late-type galaxies (typical hosts of Seyfert nuclei, 
e.g., \citealt{2003ApJ...589..774M,2004ApJ...616..707H}). 
Regarding early type galaxies, on the other hand, 
\citet{2007ApJ...655..718S} found that Seyfert nuclei preferentially accompany dusty structures. 
Thus, the presence of a nuclear dusty (equivalently gaseous) structure 
is necessary but not sufficient to trigger AGNs. 
To summarize, there is apparently no unique mechanism in Seyfert galaxies at $\ga 100$ pc from the center, 
and direct triggers of AGN activity should exist at the innermost 100 pc region. 

At that spatial scale, many studies have proposed 
that the inflowing gas would form a {\it{circumnuclear disk (CND)}}
\footnote{A CND refers to a massive gaseous disk with sizes of $\sim1-100$ parsec [pc] in this work.} 
due to the remaining angular momentum (e.g., \citealt{2005ApJ...630..167T,2008A&A...491..441V,2008ApJ...685..787B}; \citealt{2008ApJ...681...73K}). 
Indeed, such dense molecular gas disks have been found observationally around Seyfert nuclei 
(e.g., \citealt{2007A&A...468L..63K,2009ApJ...696..448H,2013ApJ...768..107H,2012A&A...537A.133D,
2012MNRAS.424.1963S,2013PASJ...65..100I,2015ApJ...811...39I,2014A&A...567A.125G}), 
although few of them have been resolved spatially. 
Recently, \citet{2013ApJ...768..107H} reported on systematic differences 
at the CND-scale between active and inactive galaxies: 
Seyfert galaxies showed more centrally concentrated profiles of 
both the stellar continuum and H$_2$ $1-0$ S(1) line emission with enhanced H$_2$ luminosity. 
Therefore, molecular surface brightness is clearly elevated in CNDs of Seyfert galaxies. 

Because a CND would be a massive reservoir of molecular gas, 
we can reasonably expect active star formation there. 
Observationally, prominent star formation has been found as a (circum-)nuclear starburst 
(e.g., \citealt{1995ApJ...452..549H,2004MNRAS.355..273C,2004ApJ...617..214I,
2007ApJ...671.1388D,2012ApJ...746..168D,2013ApJ...765...78A,2014ApJ...780...86E}). 
Interestingly, there are correlations between the star formation rate (SFR) 
and the black hole accretion rate ($\dot{M}_{\rm BH}$) in Seyfert galaxies, 
which are tighter when the SFR is measured in closer vicinity to the AGN, 
whereas it is weaker for larger-scale ($\ga \rm{kpc}$) SFR (\citealt{2012ApJ...746..168D,2014ApJ...780...86E}). 
From a theoretical viewpoint, such a correlation would indeed be predicted to 
exist due to various mechanisms (e.g., \citealt{2008ApJ...681...73K,2010MNRAS.407.1529H}). 

On the origin of the ${\rm SFR}-\dot{M}_{\rm BH}$ correlation, 
we should consider the results of \citet{2007ApJ...671.1388D}, 
who showed a time delay of $50-200$ Myr between the onset of star formation 
and the peak epoch of AGN activity (see also \citealt{2009ApJ...692L..19S,2010MNRAS.405..933W}). 
That delay might be compelling evidence of a {\it{causal}} connection between these activities: 
that star formation provides the fuel for the SMBH (Section \ref{sec5}). 
As a candidate mechanism to make such a link, \citet{2010MNRAS.407.1529H} 
suggested, for example, the importance of a series of gravitational instability and the resulting 
stellar gravitational torque based on hydro-dynamic simulations (see also \citealt{2012MNRAS.420L...8H}). 
Inside the central $r \la10$ pc, they predicted 
that a $m = 1$ mode (single-armed spiral) develops, 
which efficiently removes the angular momentum of the gas. 
Indeed, a nuclear trailing spiral (but $m = 2$) 
has been observed in the central $\sim$50 pc 
of the type 1 Seyfert galaxy NGC 1566 (\citealt{2014A&A...565A..97C}). 
However, the prevalence in other Seyfert galaxies is unclear, 
considering that there is no morphological difference between active and inactive galaxies at $\ga 100$ pc. 
More directly connected to star formation, 
mass loss from evolved stars (\citealt{1988ApJ...332..124N,1991ApJ...376..380C,1993ApJ...416...26P,2007ApJ...671.1388D}) 
or angular momentum loss due to supernova (SN)-driven turbulence 
(\citealt{2002ApJ...566L..21W,2008ApJ...681...73K,2009ApJ...702...63W,2011MNRAS.413.2633H}) 
can also enhance mass accretion. 

To determine the properties of cold molecular gas 
(e.g., mass, distribution, kinematics) is essential for further progress 
because it takes the bulk of the gas mass in the nuclear region of galaxies (e.g., \citealt{1995A&A...304....1H}). 
Such dense gas could be a source of fuel to an SMBH as well as a stellar nursery. 
Thus, it should contain information on the origin of the AGN$-$starburst connection. 
From this perspective, the pioneering work of \citet{1994ApJ...423L..27Y} showed a linear correlation 
between X-ray luminosity and CO($1-0$) luminosity in some AGNs, 
where CO($1-0$) emission lines were measured 
with single-dish telescopes (i.e., spatial resolution $>$ kpc). 
\citet{2011ApJS..195...23M} updated that result 
with improved statistics, again based on single-dish measurements. 
However, the origin of these correlations is still unclear 
as is the case for the ${\rm SFR}-\dot{M}_{\rm BH}$ correlations. 
Whether there is any dependence of the correlation strength on the probed spatial scale 
has not yet been investigated. 
Here, we note that, if we use dense gas tracers for our investigation 
rather than the conventional CO($1-0$) line that traces total molecular gas, 
including diffuse and/or foreground ones, 
we can expect less contamination, at least from the foreground component (e.g., galactic disks). 
Moreover, dense gas is indeed the source of massive star formation. 
Thus, to provide further insights on the AGN$-$starburst connection more directly 
and to understand the underlying mass accretion processes at the CND scale, 
it would be desirable to establish a correlation between AGN activity and some molecular properties 
based on emission lines that faithfully trace star forming regions, 
to check the variation in the spatial scale probed, 
and to explore the origin of the correlations. 

\subsection{This work}\label{sec1.1}
In this work, motivated by the idea above, 
we explored the possible correlation between 
$\dot{M}_{\rm BH}$ and the mass of the CND-scale molecular gas 
as a natural extension of the previous galactic scale measurements 
(\citealt{1994ApJ...423L..27Y,2011ApJS..195...23M}). 
The mass of {\it{dense}} molecular gas ($M_{\rm dense}$) was investigated 
preferentially for the reasons above. 
Thus, we compiled currently available data for a typical dense gas tracer at the millimeter band, 
specifically, the HCN($1-0$) emission line (Section \ref{sec2.1}). 
This allows us to straightforwardly compare results with 
the ${\rm SFR}-\dot{M}_{\rm BH}$ correlations, 
because $M_{\rm dense}$ is directly convertible to SFR (e.g., \citealt{2004ApJ...606..271G}). 
The detailed spatial distribution (or gas surface density) and kinematics of dense molecular gas, however, 
cannot be discussed in this paper as most CNDs have not been resolved spatially at this point. 
Such a study will be possible with future high-resolution observations 
provided by, e.g., the Atacama Large Millimeter/submillimeter Array (ALMA). 

The rest of this paper is organized as follows. 
In Section \ref{sec2}, we describe the details of the data 
and derivation of each physical parameter used. 
Section \ref{sec3} describes our regression analysis. 
The resulting correlation plots between 
HCN($1-0$) luminosity and $2-10$ keV X-ray luminosity, 
or, almost equivalently, between $M_{\rm dense}$ and $\dot{M}_{\rm BH}$, 
are presented in Section \ref{sec4}. 
Section \ref{sec5} is dedicated to a discussion 
of a SN-driven turbulence model (\citealt{2008ApJ...681...73K}) 
for demonstrative purposes, although this is not definitive currently. 
Our conclusions are summarized in Section \ref{sec6}.

\section{Data description}\label{sec2}
Here, we describe the details of the emission line, sample galaxies, 
and the procedure to estimate the physical quantities used. 
The results are listed in Tables \ref{tbl1} and \ref{tbl2}. 
Cosmology with $H_0$ = 70 km s$^{-1}$ Mpc$^{-1}$, 
$\Omega_{\rm M}$ = 0.3, $\Omega_{\rm \Lambda}$ = 0.7 was adopted throughout this work. 
We included galaxies classified as Seyfert 1.5, 1.8, and 1.9 into the Seyfert 1 category 
to maintain consistency with \citet{2014ApJ...780...86E}. 

\subsection{Line selection}\label{sec2.1}
We used the HCN($1-0$) emission line (rest frame frequency, 
$\nu_{\rm rest}$ = 88.631 GHz) as a proxy for $M_{\rm dense}$. 
The critical density for collisional excitation with H$_2$ ($n_{\rm cr}$) 
is $\sim 10^{5}$ cm$^{-3}$ in the optically thin limit. 
This line is one of the brightest ones in the 3 mm 
wavelength band in nearby galaxies after CO($1-0$). 
Unlike CO($1-0$) which traces total molecular gas 
(i.e., both dense and diffuse/foreground gas; $n_{\rm cr} \sim 10^{2-3}$ cm$^{-3}$), 
the HCN($1-0$) emission emanates selectively from dense molecular gas 
where stars are born (\citealt{2004ApJS..152...63G,2004ApJ...606..271G,2005ApJ...635L.173W}). 
Thus, using HCN($1-0$) may be more suitable physically to investigate any possible link 
between circumnuclear star formation and AGN activity. 

Indeed, molecular gas at a CND is typically so dense and warm 
($n_{\rm H_2} \ga 10^{4-5}$ cm$^{-3}$, kinetic temperature $\ga 100$ K; 
e.g., \citealt{2012A&A...537A.133D,2013PASJ...65..100I,2014A&A...570A..28V}) 
that CO molecules are readily excited to higher rotational states (e.g., \citealt{2014ApJ...795..174K}). 
However, converting the line flux of high-$J$ CO into that of CO($1-0$) 
would cause large uncertainty, unless spatially resolved information is available. 
Thus, it was better to continue to use the $J = 1-0$ transition line 
that can selectively trace dense gas, such as HCN($1-0$), to estimate $M_{\rm dense}$. 

The HCN($1-0$) line becomes more intense in AGNs 
with respect to the CO, HCO$^+$, and CS emission lines, 
compared to starburst galaxies 
(e.g., \citealt{2005AIPC..783..203K,2008ApJ...677..262K,2012A&A...537A.133D,2016ApJ...818...42I}). 
This may be a consequence of 
either an abnormal chemical composition due to AGN feedback, 
excitation (gas density, temperature), optical depth, infrared-pumping, or even a combination of them. 
However, the line optical depth of HCN($1-0$) is almost always beyond unity in most objects 
(i.e., the intensity gets more or less saturated even if HCN abundance is enhanced in AGNs). 
Thus, the potential variation in abundance 
will not severely affect our conclusions as long as 
we estimate $M_{\rm dense}$ from HCN($1-0$) line luminosity via the virial theorem. 
Using a mass conversion factor specifically estimated for AGNs (\citealt{2008ApJ...677..262K}) 
will also help to reduce this uncertainty. 

\subsection{Interferometric data}\label{sec2.2}
We compiled high-resolution (aperture $< 500$ pc) 
interferometric data for the HCN($1-0$) emission line 
from the literature and the ALMA archive to estimate 
$M_{\rm dense}$ of CNDs of nearby Seyfert galaxies. 
There is currently very little information on such high-resolution HCN($1-0$) data, 
which ultimately limits the number of our sample galaxies 
(called the {\it{interferometric sample $\equiv$ IT sample}}). 
The data were obtained with the Plateau de Bure Interferometer = PdBI 
(NGC 2273, NGC 3227, NGC 4051, and NGC 6951: \citealt{2012MNRAS.424.1963S} and references therein, 
NGC 3079 and NGC 5033: \citealt{2016MNRAS.458.1375L}), 
the Nobeyama Millimeter Array = NMA (NGC 1068: \citealt{2008Ap&SS.313..279K}), 
and ALMA (NGC 1097\footnote{ID = 2011.0.00108.S}: 
\citealt{2015A&A...573A.116M}, NGC 4579\footnote{ID = 2012.1.00456.S}: ALMA archive, 
and NGC 7469\footnote{ID = 2012.1.00165.S}: T. Izumi et al. in preparation). 
We guide a reader for the detailed data analysis to those references. 
Regarding each of the PdBI sources, the line flux/luminosity in the literature was 
measured with an aperture ($3'' - 5''$), rather than a single synthesized beam. 
This enabled us to cover the entire area of each CND in all cases. 
In NGC 1097, NGC 4579, and NGC 7469, although the line flux/luminosity 
measured with single synthesized beams placed at the AGN positions were used, 
their CNDs were not resolved with the adopted beams. 
Thus, we covered almost the entire area of each CND in these samples as well. 
For NGC 1068, there are higher resolution ($\sim$100 pc) data with PdBI (\citealt{2008JPhCS.131a2031G,2014A&A...570A..28V}). 
Nevertheless, we used the NMA data mentioned above (380 pc resolution), 
considering the relatively large CND of this galaxy ($\sim$300 pc in diameter; \citealt{2014A&A...567A.125G}). 
This enabled us to make a fair comparison among the targets 
because we could compile the line fluxes/luminosities from the entire area of each CND. 
Note that the ALMA archival data (NGC 4579) were analyzed further 
with the Common Astronomy Software Applications 
(CASA ver. 4.2.2; \citealt{2007ASPC..376..127M,2012ASPC..461..849P}). 
The details of this can be found in the Appendix-\ref{app-A}. 
We excluded galaxies showing an absorption feature in the HCN($1-0$) line from the IT sample. 
Indeed, NGC 4945 and Centaurus A (both in the ALMA archive) 
showed prominent absorption features at their nuclei. 
We also excluded NGC 5194 (M51) from the IT sample 
because it exhibited a possible feature of maser amplification 
in the HCN($1-0$) emission (\citealt{2015ApJ...799...26M}). 

Thus, we assessed the HCN($1-0$) line flux/luminosities for 
a total of 10 Seyfert galaxies (six type 1 and four type 2). 
The median distance of the samples was 17.4 Mpc. 
For each AGN from the literature, we used the line flux/luminosity listed in the reference, 
measured with an aperture sufficient to encompass the whole CND, as noted above. 
Thus, the sampled spatial scale differed among galaxies (median value $\theta_{\rm med}$ = 220 pc). 
Note that the angular sizes of the target CNDs were at most a few arcseconds (e.g., \citealt{2013PASJ...65..100I,2015ApJ...811...39I}), 
which is well below the maximum recoverable angular scale of the interferometers mentioned above. 
Consequently, the missing flux is not a major problem.

\subsubsection{Line luminosity of HCN($1-0$)}\label{sec2.2.1} 
HCN($1-0$) line luminosity $L'_{\rm HCN}$ was calculated as 
\begin{equation}\label{eq1}
\begin{split}
\left( \frac{L'_{\rm HCN}}{\rm{K ~km ~s^{-1} ~pc^2}} \right) &= 3.25 \times 10^7~ \left( \frac{S\Delta v}{\rm {Jy~km~s^{-1}}} \right) ~ \left( \frac{\nu_{\rm rest}}{\rm{GHz}} \right)^{-2} \\
&\quad \cdot \left( \frac{D_{\rm L}}{\rm{Mpc}} \right)^2 ~(1+z)^{-1}, 
\end{split}
\end{equation}
where $S\Delta v$, $D_{\rm L}$, and $z$ indicate the velocity-integrated line flux of HCN($1-0$), 
the luminosity distance to the object, and the redshift of the object, respectively (\citealt{1992ApJ...398L..29S,2005ARA&A..43..677S}). 
For the data taken from the literature, we simply used the listed $S\Delta v$ after correcting $D_{\rm L}$. 
For NGC 4579 and NGC 7469, we measured $S\Delta v$ in a zeroth moment map, 
which was computed with the CASA task \verb|immoments| without any clipping 
after subtracting the underlying continuum emission. 
The velocity ranges for the integration were chosen carefully 
to fully cover the full width at zero intensity of the HCN($1-0$) lines, 
but not to be so large as to introduce unnecessary noise. 
The RMS noises for these maps were measured in areas free of HCN($1-0$) emissions.

\subsubsection{Mass of dense molecular gas}\label{sec2.2.2}
Assuming that the HCN($1-0$) emission is emanating from an ensemble of virialized, 
non-shadowing (in space and velocity) clouds, we can estimate 
$M_{\rm dense}$ of the CND as (e.g., \citealt{1990ApJ...348L..53S}), 
\begin{equation}\label{eq2}
\begin{split}
\left( \frac{M_{\rm dense}}{M_\odot} \right) &= 2.1 ~ \left(\frac{n_{\rm H_2}}{\rm{cm^{-3}}} \right)^{1/2} ~ \left( \frac{T_{\rm b}}{\rm{K}} \right)^{-1} ~ L'_{\rm HCN} \\
&\quad \equiv X_{\rm HCN} \cdot L'_{\rm HCN},
\end{split}
\end{equation}
where $T_{\rm b}$ is the brightness temperature of the emission line. 
We direct readers to \citet{2012ApJ...751...10P} for a more sophisticated formalism 
that accounts for the departure from the assumptions above. 
Although the conversion factor ($X_{\rm HCN}$) would span a wide range, 
we adopt $X_{\rm HCN}$ = 10 $M_\odot$ (K km s$^{-1}$ pc$^2$)$^{-1}$ throughout this work. 
This factor was specifically estimated for nearby AGNs through a multi-transitional 
non-local thermodynamic equilibrium analysis 
of the single dish-based data (IRAM 30m; \citealt{2008ApJ...677..262K}). 
We adopted 0.30 dex uncertainty for this factor, 
as estimated by the authors above, 
which dominates the total uncertainty of $M_{\rm dense}$.

\subsubsection{Black hole mass}\label{sec2.2.3}
$M_{\rm BH}$ is used to discuss AGN properties (Section \ref{sec2.4}) 
and mass accretion mechanisms (Section \ref{sec5}). 
We favored values derived from stellar 
or gas kinematics (NGC 1097 and NGC 3227), 
Very Long Baseline Interferometry (VLBI) maser 
observations (NGC 1068 and NGC 2273), 
and reverberation mapping (NGC 4051 and NGC 7469). 
We fixed a virial factor of $f = 4.3$ for the reverberation method 
(\citealt{2013ApJ...773...90G,2015PASP..127...67B}\footnote{\url{http://www.astro.gsu.edu/AGNmass/}}). 
For the other galaxies, the $\sigma_*$ compiled in HyperLeda\footnote{\url{http://leda.univ-lyon1.fr}} 
(\citealt{2014A&A...570A..13M}) were applied to the $M_{\rm BH}-\sigma_*$ relationship 
constructed by \citet{2009ApJ...698..198G}. 
An intrinsic scatter (0.44 dex) accompanying the relation was also taken into account 
when estimating the total uncertainty in $M_{\rm BH}$.

\subsubsection{Bolometric luminosity and mass accretion rate}\label{sec2.2.4}
An absorption-corrected $2-10$ keV hard X-ray luminosity 
($L_{\rm 2-10 keV}$) catalogued in published papers 
was considered a proxy for the bolometric luminosity of the AGN ($L_{\rm Bol}$; e.g., \citealt{2008ARA&A..46..475H}). 
The bolometric correction of \citet{2004MNRAS.351..169M} 
was applied to the $L_{\rm 2-10 keV}$. 
The uncertainties in the $L_{\rm Bol}$ would be mainly 
driven by the scatter on the correction and the time variability of the $L_{\rm 2-10keV}$, 
which, generally, are significantly larger than 
the statistical error in the $L_{\rm 2-10 keV}$. 
We adopted 0.30 dex uncertainty for each, which is likely to be sufficient. 
Adding these in quadrature, 0.42 dex 
uncertainty was set for $L_{\rm Bol}$. 

The mass accretion rate onto an SMBH was subsequently estimated using the following relationship (\citealt{2012NewAR..56...93A}), 
\begin{equation}\label{eq3}
\left( \frac{\dot{M}_{\rm BH}}{M_\odot ~\rm{yr^{-1}}} \right) = 0.15 \left( \frac{0.1}{\eta} \right) \left( \frac{L_{\rm Bol}}{\rm{10^{45} ~erg ~s^{-1}}} \right)
\end{equation}
where $\eta = 0.1$ is a typical value for mass-to-energy conversion efficiency 
in the local universe (\citealt{2004MNRAS.351..169M}). 
Any uncertainty accompanying this relationship was not taken into account. 
Thus, the total uncertainty of $\dot{M}_{\rm BH}$ 
is the same as that of $L_{\rm Bol}$, i.e., 0.42 dex. 
Note that models predict a drop of the $\eta$ 
in a low accretion phase (\citealt{1995ApJ...452..710N}) 
due to advection of matter in an accretion disk. 
In that case, the estimated value above 
would be the lower limit of $\dot{M}_{\rm BH}$. 
We ignore this potential influence here, 
but that may be a subject of future studies. 

Regarding NGC 6951, we could not find 
an absorption-corrected $L_{\rm 2-10keV}$ in the literature. 
However, rather than discarding this valuable sample 
with high-resolution HCN($1-0$) measurements, 
we used another proxy for $L_{\rm Bol}$, 
specifically an [\ion{O}{4}] line luminosity ($L_{\rm [O IV]}$). 
The line flux collected by 
\citet{2009ApJ...698..623D} and the bolometric correction of 
$L_{\rm Bol}$ = 810 $\times$ $L_{\rm [O IV]}$ 
(for type 2 Seyfert galaxies; \citealt{2009ApJ...700.1878R}) were applied. 
The uncertainty in this $L_{\rm Bol}$ was assumed to 
be at the same level as that derived from 
$L_{\rm 2-10keV}$ for simplicity: i.e., 0.42 dex. 

\subsection{Single-dish data}\label{sec2.3}
To investigate the impact of an aperture 
for sampling $L'_{\rm HCN}$ in this study (Section \ref{sec3}), 
we compiled HCN($1-0$) flux data for nearby Seyfert galaxies 
obtained with single-dish telescopes from the literature. 
Again, the number of HCN($1-0$) detections limited the total number 
of our sample (the {\it{single-dish sample $\equiv$ SD sample}}). 
We excluded merging galaxies from our sample, because we could not judge 
from which galaxy (or both) the line emission came, 
considering the coarse spatial resolution. 
The angular resolutions of the telescopes were, 
29$\arcsec$ (IRAM 30 m), 57$\arcsec$ (SEST), 
44$\arcsec$ (OSO), 72$\arcsec$ (NRAO 12 m), 
62$\arcsec$ (FCRAO 14 m), and 19$\arcsec$ (NRO 45 m), respectively, 
at the $\nu_{\rm rest}$ of HCN($1-0$). 
These resolutions typically correspond to more than a few kpc 
and sometimes even reach $>10$ kpc at the distance of the sample galaxies. 
This indicates that these observations traced dense molecular gas 
existing over a bulge-scale (typically a few kpc) or larger (i.e., an entire galaxy). 
We should also emphasize that the spatial resolution of 
$\sim$ a few kpc would be insufficient to separate a CND (100 pc scale) 
from other components in that central region, 
such as a kpc-scale circumnuclear starburst ring (e.g., \citealt{2013PASJ...65..100I,2015ApJ...811...39I}). 

We adopted the same procedure described in Section \ref{sec2.2} 
to achieve $M_{\rm dense}$, $M_{\rm BH}$, $L_{\rm Bol}$, and $\dot{M}_{\rm BH}$. 
In most cases, we had no choice but 
to use the $M_{\rm BH}-\sigma_*$ relationship 
(\citealt{2009ApJ...698..198G}) to estimate $M_{\rm BH}$. 
This $\sigma_*$ was collected mainly from HyperLeda, and otherwise from the literature. 
A total of 32 samples (12 type 1 and 20 type 2) were assessed. 
Their median distances and spatial resolutions 
were 26.5 Mpc and 5.5 kpc, respectively.

\subsection{Comments on the AGN sample}\label{sec2.4}
We first note that neither the IT nor SD sample is homogeneous, 
because they were simply complied from the literature or the archive. 
Around 75\% and 60\% of our IT and SD samples, respectively, 
belong to the revised Shapley-Ames (RSA) catalog (\citealt{1987rsac.book.....S,1995ApJ...454...95M}), 
which includes Seyfert galaxies brighter than $B_{\rm T} = 13$ mag. 
The RSA sample is a magnitude-limited one, sensitive to low-luminosity AGNs. 
However, as shown in Table \ref{tbl1}, the morphology of our sample 
was mostly the barred spiral (SAB or SB type), 
which clearly contrast against the broader distribution 
of the whole RSA Seyfert sample. 

\begin{figure}
\epsscale{1}
\plotone{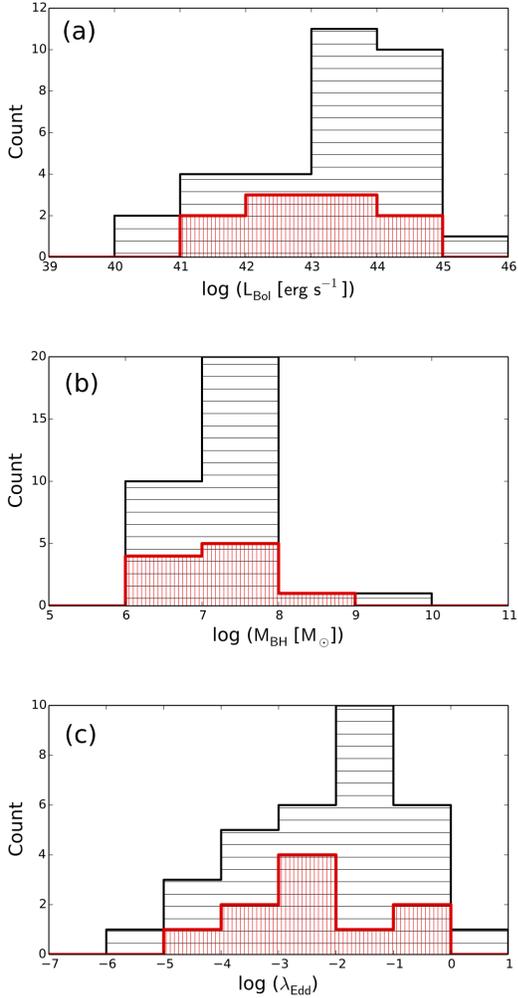}
\caption{
Distributions of (a) bolometric luminosity $L_{\rm Bol}$, (b) black hole mass $M_{\rm BH}$, 
and (c) Eddington ratio $\lambda_{\rm Edd}$ of the sample galaxies. 
Data obtained with interferometers (IT sample: red, vertical line) 
and single-dish telescopes (SD sample: black, horizontal line) are shown. 
}
\label{figure1}
\end{figure}

Figure \ref{figure1} shows the distribution of the $L_{\rm Bol}$, 
$M_{\rm BH}$, and $\lambda_{\rm Edd}$ of our sample AGNs. 
Here, $\lambda_{\rm Edd}$ denotes the Eddington ratio in the luminosity regime
\footnote{$\lambda_{\rm Edd}$ = $L_{\rm Bol}$/$L_{\rm Edd}$, 
where $L_{\rm Edd}$ [erg s$^{-1}$] = 1.26 $\times$ 10$^{38}$ $M_{\rm BH}$ [$M_\odot$].}. 
The logarithmic median values of 
($L_{\rm Bol}$, $M_{\rm BH}$, $\lambda_{\rm Edd}$) are 
(42.80, 7.14, -2.43) and (43.41, 7.42, -2.08) 
for the IT and SD samples, respectively. 
The resulting $\lambda_{\rm Edd}$ is distributed 
over a wide range: i.e., $\sim$ 5 orders of magnitude. 
This might suggest the limited applicability of Equation (\ref{eq3}) 
for all of our samples under the fixed value of $\eta$ = 0.1, 
because it has been predicted that the accretion disk will change its state 
at $\lambda_{\rm Edd} \sim 10^{-3}$ (e.g., \citealt{2013LRR....16....1A}). 
We nevertheless stick to the current estimation of $\dot{M}_{\rm BH}$ 
because the actual $\eta$ is not observationally 
constrained for low-$\lambda_{\rm Edd}$ objects. 

\begin{figure}
\epsscale{1}
\plotone{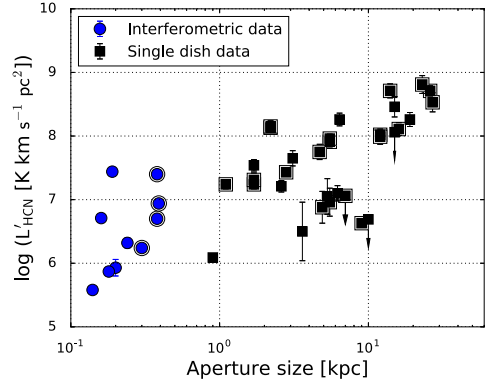}
\caption{
Scatter plot of the spatial resolution and HCN($1-0$) line luminosity 
including both the IT sample (blue circle) and the SD sample (black square). 
There is a positive correlation between these two quantities 
(correlation coefficient = 0.71$^{+0.07}_{-0.08}$). 
The doubled symbols denote type 2 AGNs hereafter. 
}
\label{figure2}
\end{figure}

Based on these data, the SD sample seems to be slightly biased 
towards higher-luminosity objects than the IT sample, 
likely because there are few high-luminosity (or high $\lambda_{\rm Edd}$) 
AGNs in the very nearby universe (e.g., $D_{\rm L} \la 20$ Mpc), 
where the CND could be spatially well separated from, for example, the surrounding starburst ring, 
at the HCN($1-0$) emission line in past interferometric observations. 
However, we could not clearly reject the null hypothesis that 
both the IT and the SD samples are drawn from the same distribution. 
Indeed, the Kolmogorov$-$Smirnov (KS) test of the IT and SD distributions shown in Figure \ref{figure1} 
returned marginal (or even high) $p$-values, which were (0.33, 0.75, 0.75) for 
($L_{\rm Bol}$, $M_{\rm BH}$, $\lambda_{\rm Edd}$), respectively. 
Thus, admitting the poor statistics of the sample and taking 
the relatively large uncertainty for each parameter into account, 
we ignored any (possible) systematic difference in 
the intrinsic AGN properties between the two sample groups. 
In this way, the only difference between the two sample groups 
that influences our study is $M_{\rm dense}$, 
which directly reflects the orders-of-magnitude 
different spatial scales for HCN($1-0$) measurements (Figure \ref{figure2}): 
our whole sample exhibits a positive correlation between the spatial resolution 
and the HCN($1-0$) line luminosity with a correlation coefficient of 0.71$^{+0.07}_{-0.08}$ 
(excluding the upper limits on $L'_{\rm HCN}$; see Section \ref{sec3} for details of the correlation analysis).

\begin{table*}
\begin{center}
\caption{Properties of the sample galaxies \label{tbl1}}
\begin{tabular}{ccccccccccc}
\tableline\tableline
Target & $D_{\rm L}$ & Morphology & AGN type & log ($L_{\rm 2-10 keV}$) & Ref. 1 & log ($L'_{\rm HCN}$) & $\theta_{\rm res}$ & Ref. 2 & log ($M_{\rm BH}$) & Ref. 3 \\
(1) & (2) & (3) & (4) & (5) & (6) & (7) & (8) & (9) & (10) & (11) \\ \hline \hline
\multicolumn{11}{c}{Interferometric data (IT-sample)} \\
\tableline
NGC 1068 & 16.3 & (R)SA(rs)b & 2 & 43.02$\pm$0.30 & (1) & 7.40$\pm$0.07 & 0.38 & (7) & 6.96$\pm$0.02 & (24), maser \\
NGC 1097 & 18.2 & SB(s)b & 1 & 40.84$\pm$0.30 & (2) & 6.71$\pm$0.06 & 0.16 & (8) & 8.15$\pm$0.10 & (25), gas \\
NGC 2273 & 26.4 & SB(r)a & 2 & 42.73$\pm$0.30 & (1) & 6.70$\pm$0.05 & 0.38$^\sharp$ & (9) & 6.89$\pm$0.02 & (26), maser \\
NGC 3079 & 16.0 & SB(s)c edge-on & 2 & 42.02$\pm$0.30 & (1) & 6.94$\pm$0.04 & 0.39$^\sharp$ & (10) & 7.90$\pm$0.63 & (27,28), $M_{\rm BH}-\sigma_*$ \\
NGC 3227 & 16.6 & SAB(s)a pec & 1.5 & 42.07$\pm$0.30 & (2) & 6.32$\pm$0.08 & 0.24$^\sharp$ & (9) & 7.18$\pm$0.30 & (29), stellar \\
NGC 4051 & 10.0 & SAB(rs)bc & 1.5 & 41.13$\pm$0.30 & (2) & 5.58$\pm$0.05 & 0.14$^\sharp$ & (9) & 6.13$\pm$0.14 & (30), reverberation \\
NGC 4579 & 21.8 & SAB(rs)b & 1.5 & 41.33$\pm$0.30 & (2) & 5.93$\pm$0.13 & 0.20 & (11) & 7.76$\pm$0.45 & (27,28), $M_{\rm BH}-\sigma_*$ \\
NGC 5033 & 12.5 & SA(s)c & 1.9 & 40.91$\pm$0.30 & (2) & 5.87$\pm$0.05 & 0.18$^\sharp$ & (10) & 7.38$\pm$0.46 & (27,28), $M_{\rm BH}-\sigma_*$ \\
NGC 6951 & 20.4 & SAB(rs)bc & 2 & -$^\dag$ & -$^\dag$ & 6.24$\pm$0.04 & 0.30$^\sharp$ & (9) & 7.10$\pm$0.50 & (27,28), $M_{\rm BH}-\sigma_*$ \\
NGC 7469 & 70.8 & (R')SAB(rs)a & 1.2 & 43.17$\pm$0.30 & (2) & 7.44$\pm$0.04 & 0.19 & (12) & 6.97$\pm$0.05 & (30), reverberation \\ \hline \hline
\multicolumn{11}{c}{Single dish data (SD-sample)} \\ \hline
Cen A & 7.8 & S0 pec & 2 & 41.90$\pm$0.30 & (3) & 7.24$\pm$0.07 & 1.1 & (13) & 7.65$\pm$0.04 & (31), gas \\
Circinus & 6.2 & SA(s)b & 2 & 42.62$\pm$0.30 & (1) & 7.24$\pm$0.07 & 1.7 & (14) & 6.23$\pm$0.08 & (32), maser \\
IRAS 05189-2524 & 188.2 & - & 2 & 44.20$\pm$0.30 & (4) & 8.71$\pm$0.08 & 26 & (15) & 7.42$\pm$0.40 & (27,28), $M_{\rm BH}-\sigma_*$ \\ 
Mrk 273 & 166.5 & pec & 2 & 42.87$\pm$0.30 & (2) & 8.81$\pm$0.14 & 23 & (16) & 9.17$\pm$0.05 & (33), maser \\
Mrk 331 & 80.3 & Sa & 2? & 40.70$\pm$0.30 & (4) & 8.53$\pm$0.15 & 27 & (17) & 6.81$\pm$0.47 & (27,28), $M_{\rm BH}-\sigma_*$ \\ 
NGC 34 & 85.3 & pec & 2 & 42.00$\pm$0.30 & (4) & 8.02$\pm$0.07 & 12 & (15) & 7.71$\pm$0.52 & (27,28), $M_{\rm BH}-\sigma_*$ \\
NGC 660 & 12.2 & SB(s)a pec & 2 & 39.40$\pm$0.30 & (4) & 7.31$\pm$0.11 & 1.7 & (18) & 7.12$\pm$0.47 & (27,28), $M_{\rm BH}-\sigma_*$ \\
NGC 931 & 72.2 & SAbc & 1 & 43.29$\pm$0.30 & (2)& $<$8.06 & 15 & (19) & 7.54$\pm$0.60 & (27,28), $M_{\rm BH}-\sigma_*$ \\
NGC 1068 & 16.3 & (R)SA(rs)b & 2 & 43.02$\pm$0.30 & (1) & 8.13$\pm$0.02 & 2.2 & (20) & 6.96$\pm$0.02 & (24), maser \\
NGC 1097 & 18.2 & SB(s)b & 1 & 40.84$\pm$0.30 & (2) & 7.54$\pm$0.09 & 1.7 & (21) & 8.15$\pm$0.10 & (25), gas \\
NGC 1365 & 23.5 & SB(s)b & 1.8 & 42.25$\pm$0.30 & (2) & 8.26$\pm$0.10 & 6.4 & (19) & 7.50$\pm$0.51 & (27,28), $M_{\rm BH}-\sigma_*$ \\
NGC 1667 & 65.7 & SAB(r)c & 2 & 42.37$\pm$0.30 & (5) & 8.71$\pm$0.11 & 14 & (19) & 7.82$\pm$0.45 & (27,28), $M_{\rm BH}-\sigma_*$ \\
NGC 2273 & 26.4 & SB(r)a & 2 & 42.73$\pm$0.30 & (1) & 6.96$\pm$0.22 & 5.5 & (19) & 6.89$\pm$0.02 & (26), maser \\
NGC 3079 & 16.0 & SB(s)c edge-on & 2 & 42.02$\pm$0.30 & (1) & 7.91$\pm$0.10 & 5.5 & (17) & 7.90$\pm$0.63 & (27,28), $M_{\rm BH}-\sigma_*$ \\
NGC 3147 & 40.3 & SA(rs)bc & 2 & 41.40$\pm$0.30 & (4) & 7.96$\pm$0.07 & 5.5 & (22) & 7.42$\pm$0.50 & (27,28), $M_{\rm BH}-\sigma_*$ \\
NGC 4258 & 6.4 & SAB(s)bc & 1.9 & 40.57$\pm$0.30 & (2) & 6.09$\pm$0.04 & 0.9 & (16) & 7.56$\pm$0.03 & (34), maser \\
NGC 4388 & 36.3 & SA(s)b edge-on & 2 & 43.20$\pm$0.30 & (2) & 6.88$\pm$0.25 & 4.9 & (16) & 6.92$\pm$0.01 & (26), maser \\
NGC 4593 & 38.8 & (R)SB(rs)b & 1 & 42.80$\pm$0.30 & (2) & 7.05$\pm$0.28 & 5.3 & (16) & 6.89$\pm$0.09 & (30), reverberation \\
NGC 4945 & 8.6 & SB(s)cd edge-on & 2 & 42.22$\pm$0.30 & (6) & 8.15$\pm$0.09 & 2.2 & (23) & 6.46$\pm$0.03 & (35), maser \\
NGC 5005 & 13.5 & SAB(rs)bc & 2 & 40.02$\pm$0.30 & (2) & 7.75$\pm$0.12 & 4.7 & (17) & 7.84$\pm$0.45 & (27,28), $M_{\rm BH}-\sigma_*$ \\
NGC 5033 & 12.5 & SA(s)c & 1.9 & 40.91$\pm$0.30 & (2) & 7.21$\pm$0.09 & 2.6 & (19) & 7.38$\pm$0.46 & (27,28), $M_{\rm BH}-\sigma_*$ \\
NGC 5135 & 59.3 & SB(s)ab & 2 & 43.10$\pm$0.30 & (1) & 8.11$\pm$0.10 & 16 & (19) & 7.24$\pm$0.46 & (27,36), $M_{\rm BH}-\sigma_*$ \\
NGC 5194 & 6.6 & SA(s)bc pec & 2 & 41.54$\pm$0.30 & (1) & 6.63$\pm$0.02 & 9.0 & (20) & 6.60$\pm$0.49 & (27,28), $M_{\rm BH}-\sigma_*$ \\
NGC 5347 & 33.6 & (R')SB(rs)ab & 2 & 42.39$\pm$0.30 & (1) & $<$7.06 & 7.0 & (19) & 6.73$\pm$0.55 & (27,28), $M_{\rm BH}-\sigma_*$ \\
NGC 5506 & 26.6 & Sa pec edge-on & 1.9 & 43.01$\pm$0.30 & (2) & 6.50$\pm$0.46 & 3.6 & (16) & 7.87$\pm$0.49 & (27,28), $M_{\rm BH}-\sigma_*$ \\
NGC 5548 & 74.5 & (R')SA(s)0/a & 1.2 & 43.42$\pm$0.30 & (2) & $<$6.69 & 10 & (16) & 7.73$\pm$0.12 & (30), reverberation \\
NGC 6814 & 22.4 & SAB(rs)bc & 1.5 & 42.18$\pm$0.30 & (2) & 7.10$\pm$0.12 & 6.2 & (19) & 7.05$\pm$0.06 & (30), reverberation \\
NGC 6951 & 20.4 & SAB(rs)bc & 2 & -$^\dag$ & -$^\dag$ & 7.43$\pm$0.01 & 2.8 & (20) & 7.10$\pm$0.50 & (27,28), $M_{\rm BH}-\sigma_*$ \\
NGC 7130 & 70.0 & Sa pec & 1.9 & 43.10$\pm$0.30 & (1) & 8.26$\pm$0.11 & 19 & (19) & 7.48$\pm$0.46 & (27,36), $M_{\rm BH}-\sigma_*$ \\
NGC 7469 & 70.8 & (R')SAB(rs)a & 1.2 & 43.17$\pm$0.30 & (2) & 8.46$\pm$0.16 & 15 & (19) & 6.97$\pm$0.05 & (30), reverberation \\
NGC 7479 & 34.2 & SB(s)c & 2 & 41.17$\pm$0.30 & (5) & 7.99$\pm$0.12 & 12 & (17) & 7.61$\pm$0.46 & (27,28), $M_{\rm BH}-\sigma_*$ \\
NGC 7582 & 22.6 & (R')SB(s)ab & 1 & 42.60$\pm$0.30 & (4) & 7.65$\pm$0.12 & 3.1 & (18) & 7.56$\pm$0.51 & (27,28), $M_{\rm BH}-\sigma_*$ \\
\tableline
\end{tabular}
\tablecomments{Column 1: name of the sample galaxy. 
Column 2: luminosity distance to the object in [Mpc]. 
Column 3: morphology of the host galaxy recorded in the NED (\url{http://ned.ipac.caltech.edu}). 
Mrk 331 is classified to Sa based on the record in HyperLeda (no record in the NED). 
Column 4: AGN type. We follow the classification of 
\citet{1995ApJ...454...95M} for the RSA Seyfert samples. 
For the non-RSA Seyfert samples, we mostly adopt the classification by the NED. 
The type of Mrk 331 is not clear, then we follow the one in \citet{2004ApJS..152...63G}. 
In order to keep a consistency with \citet{2014ApJ...780...86E}, 
we re-categorize type 1.5, 1.8, and 1.9 into type 1 in the subsequent plots. 
Columns 5 and 6: the logarithmic scale value of 
absorption corrected $2-10$ keV luminosity in [erg s$^{-1}$] and its reference. 
Zero padding is applied to some data. 
This quantity is used to estimate $L_{\rm Bol}$ of the AGN. 
$^\dag$We used [\ion{O}{4}] line luminosity 
to estimate $L_{\rm Bol}$ (\citealt{2009ApJ...700.1878R,2009ApJ...698..623D}). 
Columns 7, 8, and 9: the logarithmic scale 
value of HCN($1-0$) line luminosity in [K km s$^{-1}$ pc$^2$], 
the aperture used to measure the luminosity in [kpc], and the reference for them. 
$^\sharp$For NGC 2273, NGC 3227, NGC 4051, and NGC 6951 in the IT-sample, 
we used the line fluxes measured for the central 
3 arcsec region as reported in \citet{2012MNRAS.424.1963S}. 
Similarly, for NGC 3079 and NGC 5033, 5 arcsec and 3 arcsec apertures are used as shown in \citet{2016MNRAS.458.1375L}. 
Systematic uncertainty is included. 
We assume 10\% and 15\% for the statistical and systematic uncertainties unless mentioned in the reference. 
Columns 10 and 11: the logarithmic scale value of black hole mass in [$M_\odot$] with the reference and the method for the estimation. 
References: (1) \citet{2012ApJ...748..130M}, 
(2) \citet{2014ApJ...783..106L}, 
(3) \citet{2012MNRAS.420.2087D},
(4) \citet{2011MNRAS.413.1206B}, 
(5) \citet{2006A&A...455..173P}, 
(6) \citet{2004A&A...418..465L}, 
(7) \citet{2008Ap&SS.313..279K}, 
(8) \citet{2015A&A...573A.116M}, 
(9) \citet{2012MNRAS.424.1963S}, 
(10) \citet{2016MNRAS.458.1375L}, 
(11) from ALMA archive (this work), 
(12) Izumi et al. in preparation, 
(13) \citet{2000A&A...359..483W}, 
(14) \citet{2001A&A...367..457C}, 
(15) \citet{2015ApJ...814...39P}, 
(16) \citet{2011MNRAS.418.1753J}, 
(17) \citet{2004ApJS..152...63G}, 
(18) \citet{2008A&A...477..747B}, 
(19) \citet{2000A&AS..141..193C}, 
(20) \citet{2008ApJ...677..262K}, 
(21) \citet{2003PASJ...55L...1K}, 
(22) \citet{1992ApJ...387L..55S}, 
(23) \citet{2004A&A...422..883W}, 
(24) \citet{2003A&A...398..517L}, 
(25) \citet{2015ApJ...806...39O}, 
(26) \citet{2011ApJ...727...20K}, 
(27) \citet{2009ApJ...698..198G}, 
(28) HyperLeda, 
(29) \citet{2006ApJ...646..754D}, 
(30) \citet{2015PASP..127...67B} and references therein, 
(31) \citet{2010PASA...27..449N}, 
(32) \citet{2003ApJ...590..162G}, 
(33) \citet{2004A&A...419..887K}, 
(34) \citealt{1995Natur.373..127M}, 
(35) \citet{1997ApJ...481L..23G}, 
(36) \citet{2005MNRAS.359..765G} 
}
\end{center}
\end{table*}

\begin{table}
\begin{center}
\caption{$M_{\rm dense}$ and $\dot{M}_{\rm BH}$ of the sample galaxies \label{tbl2}}
\begin{tabular}{ccc}
\tableline\tableline
Target & log ($M_{\rm dense}$) & log ($\dot{M}_{\rm BH}$) \\
(1) & (2) & (3) \\ 
\hline \hline
\multicolumn{3}{c}{Interferometric data (IT-sample)} \\  \hline
NGC 1068 & 8.40$\pm$0.31 &  -1.55$\pm$0.42\\
NGC 1097 & 7.71$\pm$0.30 & -4.10$\pm$0.42 \\
NGC 2273 & 7.70$\pm$0.30 & -1.91$\pm$0.42 \\
NGC 3079 & 7.94$\pm$0.30 & -2.76$\pm$0.42 \\
NGC 3227 & 7.32$\pm$0.31 & -2.70$\pm$0.42 \\
NGC 4051 & 6.58$\pm$0.30 & -3.78$\pm$0.42 \\
NGC 4579 & 6.93$\pm$0.33 & -3.56$\pm$0.42 \\
NGC 5033 & 6.87$\pm$0.30 & -4.02$\pm$0.42 \\
NGC 6951 & 7.24$\pm$0.30 & -3.29$\pm$0.42 \\
NGC 7469 & 8.44$\pm$0.30 & -1.36$\pm$0.42 \\ \hline \hline
\multicolumn{3}{c}{Single Dish data (SD-sample)} \\ \hline
Cen A & 8.24$\pm$0.31 & -2.91$\pm$0.42 \\
Circinus & 8.24$\pm$0.31 & -2.05$\pm$0.42 \\
IRAS 05189-2524 & 9.71$\pm$0.31 & -0.03$\pm$0.42 \\
Mrk 273 & 10.27$\pm$0.32 & -1.74$\pm$0.42 \\
Mrk 331 & 9.53$\pm$0.34 & -4.25$\pm$0.42 \\
NGC 34 & 9.02$\pm$0.31 & -2.79$\pm$0.42 \\
NGC 660 & 8.31$\pm$0.32 & -5.59$\pm$0.42 \\
NGC 931 & $<$9.06 & -1.21$\pm$0.42 \\
NGC 1068 & 9.13$\pm$0.30 & -1.55$\pm$0.42 \\
NGC 1097 & 8.54$\pm$0.31 & -4.10$\pm$0.42 \\
NGC 1365 & 9.26$\pm$0.32 & -2.48$\pm$0.42 \\
NGC 1667 & 9.71$\pm$0.32 & -2.34$\pm$0.42 \\
NGC 2273 & 7.96$\pm$0.37 & -1.91$\pm$0.42 \\
NGC 3079 & 8.91$\pm$0.32 & -2.76$\pm$0.42 \\
NGC 3147 & 8.96$\pm$0.31 & -3.48$\pm$0.42 \\
NGC 4258 & 7.09$\pm$0.30 & -4.39$\pm$0.42 \\
NGC 4388 & 7.88$\pm$0.39 & -1.32$\pm$0.42 \\
NGC 4593 & 8.05$\pm$0.41 & -1.82$\pm$0.42 \\
NGC 4945 & 9.15$\pm$0.31 & -2.52$\pm$0.42 \\
NGC 5005 & 8.75$\pm$0.32 & -4.96$\pm$0.42 \\
NGC 5033 & 8.21$\pm$0.31 & -4.02$\pm$0.42 \\
NGC 5135 & 9.11$\pm$0.32 & -1.45$\pm$0.42 \\
NGC 5194 & 7.63$\pm$0.30 & -2.50$\pm$0.42 \\
NGC 5347 & $<$8.06 & -2.32$\pm$0.42 \\
NGC 5506 & 7.50$\pm$0.55 & -1.56$\pm$0.42 \\
NGC 5548 & $<$7.69 & -1.04$\pm$0.42 \\
NGC 6814 & 8.10$\pm$0.32 & -2.58$\pm$0.42 \\
NGC 6951 & 8.43$\pm$0.30 & -3.29$\pm$0.42 \\
NGC 7130 & 9.26$\pm$0.32 & -1.45$\pm$0.42 \\
NGC 7469 & 9.46$\pm$0.34 & -1.36$\pm$0.42 \\
NGC 7479 & 8.99$\pm$0.32 & -3.73$\pm$0.42 \\
NGC 7582 & 8.65$\pm$0.32 & -2.07$\pm$0.42 \\
\tableline
\end{tabular}
\tablecomments{
Column 1: name of the target galaxy. 
Columns 2 and 3: logarithmic values of $M_{\rm dense}$ [$M_\odot$] and $\dot{M}_{\rm BH}$ [$M_\odot$ yr$^{-1}$]. 
Numbers in Table \ref{tbl1} are used to estimate these values 
(see the derivation in Section \ref{sec2}).
}
\end{center}
\end{table}

\section{Regression analysis}\label{sec3}
\begin{figure*}
\epsscale{1.2}
\plotone{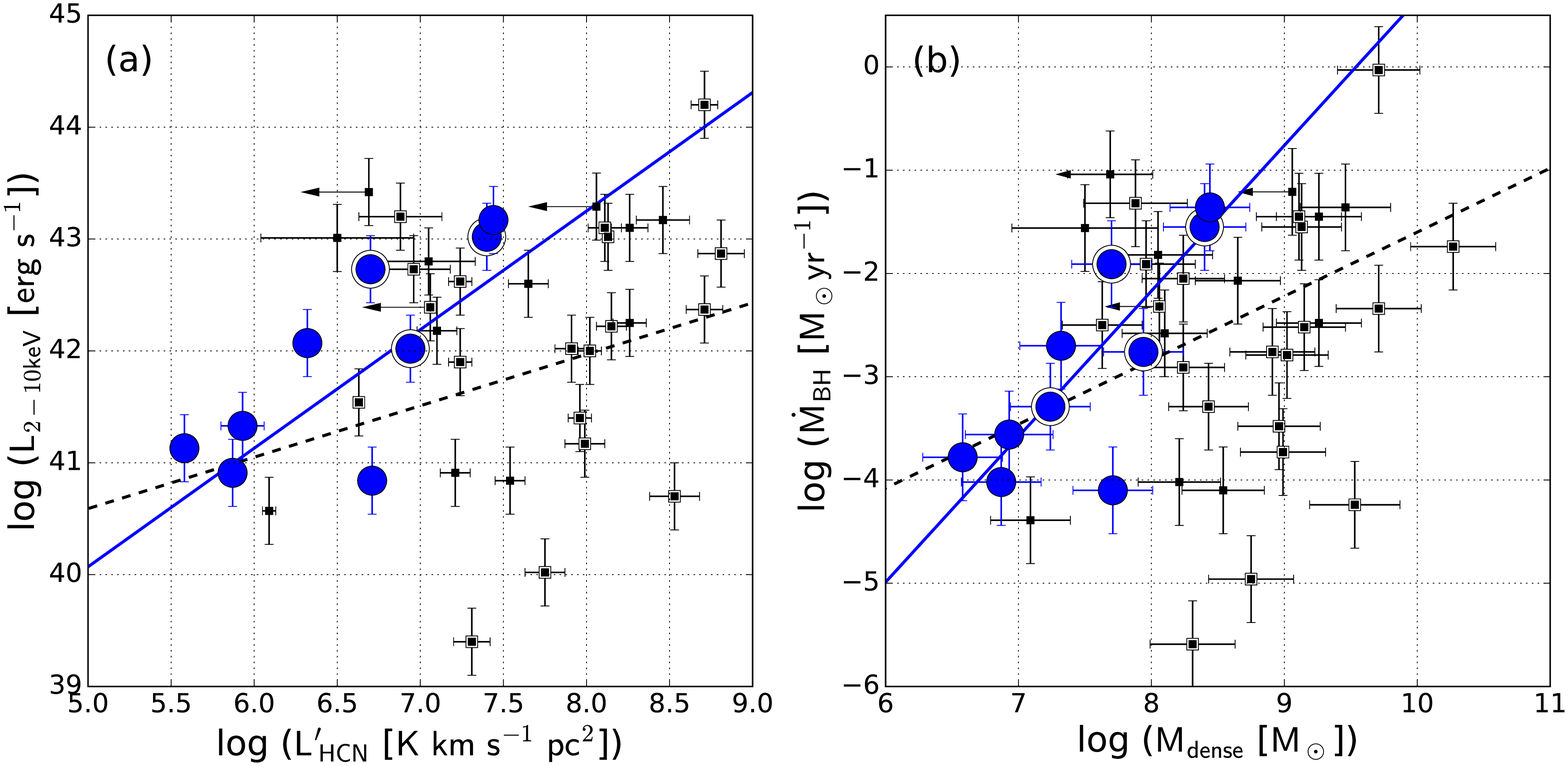}
\caption{
Observed scatter plot of (a) $L'_{\rm HCN}-L_{\rm 2-10keV}$ and (b) $M_{\rm dense}-\dot{M}_{\rm BH}$, on a logarithmic scale. 
Blue circles and black squares indicate that the HCN($1-0$) emission was obtained with 
interferometers (median aperture $\theta_{\rm med} = 220$ pc; IT sample) 
and single-dish telescopes ($\theta_{\rm med} = 5.5$ kpc; SD sample), respectively. 
The best-fit regression lines are shown by the blue solid and the black dashed line for the IT and SD samples, respectively. 
See also Section \ref{sec2} for the derivation of each parameter and Section \ref{sec3} for the details of the regression analysis. 
The best-fit regression parameters can be found in Table \ref{tbl3}. 
Note that the interferometric data of NGC 6951 was not plotted in (a) 
because of the lack of absorption-corrected $2-10$ keV X-ray luminosity in the literature. 
}
\label{figure3}
\end{figure*}

Figure \ref{figure3} shows the observed scatter plots 
of the quantities in Tables \ref{tbl1} and \ref{tbl2}, 
indicating the maximum number of data points currently available. 
Excluding those with upper limits, the median values of 
($L'_{\rm HCN}$, $L_{\rm 2-10keV}$, $M_{\rm dense}$, $\dot{M}_{\rm BH}$) 
were (6.51, 42.02, 7.51, -3.03) and (7.75, 42.37, 8.75, -2.41) 
for the IT and SD samples, respectively, on a logarithmic scale. 
Regarding the IT sample, one might see a trend 
of positive correlations in both plots, indicating the importance of CNDs 
as an external driver of AGN activity. 
To study such a view quantitatively in more detail, 
we applied the linear regression method 
developed by \citet{2007ApJ...665.1489K} 
for the two sets of variables: ($L'_{\rm HCN}$, $L_{\rm 2-10keV}$) 
and ($M_{\rm dense}$, $\dot{M}_{\rm BH}$). 
Because we used the same $X_{\rm HCN}$ 
and the fixed $\eta$ in Equation (\ref{eq3}) for all sample galaxies, 
and the conversion factors from $L_{\rm 2-10keV}$ 
to $L_{\rm Bol}$ (\citealt{2004MNRAS.351..169M}) 
are comparable in the luminosity range of our sample, 
these two sets produce virtually equivalent correlations. 
However, we should also note that any uncertainty in $\eta$ and the conversion from $L_{\rm Bol}$ to $\dot{M}_{\rm BH}$ 
were not taken into account, which could have influenced our results. 
Nevertheless, we argue that our work is an important step 
to better understand the physical link between the content of circumnuclear molecular gas and the AGN event. 
Further progress from the theoretical and the observational sides is required to refine these data. 

The procedure for the analysis is available 
from the IDL Astronomy User's Library
\footnote{\url{http://idlastro.gsfc.nasa.gov}} as \verb|linmix_err|. 
This is a Bayesian-based algorithm that can handle errors on both axes, 
upper limits on the dependent variable, and the intrinsic scatter. 
It also returns a linear correlation strength. 
The assumed formula for the regression is 
\begin{equation}\label{eq4}
\log \zeta_i = \alpha + \beta \times \log \xi_i + \epsilon_i, 
\end{equation}
where $\alpha$, $\beta$, and $\epsilon_i$ are the intercept, slope, 
and intrinsic scatter of a two-dimensional 
regression line for variables ($\xi$, $\zeta$), 
and $i$ labels each object. 
Here, $\epsilon_i$ is assumed to follow a Gaussian distribution 
with a mean zero and a constant variance of $\sigma^2_{\rm int}$ 
(we show values of $\sigma_{\rm int}$ as the intrinsic scatter). 
However, each measurement for the variable ($\xi$, $\zeta$) can be expressed as 
($x_i$, $y_i$), which have random measurement errors ($\epsilon_{x,i}$, $\epsilon_{y,i}$). 
Thus, $x_i = \xi_i + \epsilon_{x,i}$ and $y_i = \zeta_i + \epsilon_{y,i}$. 

The \verb|linmix_err| procedure uses a Markov-chain Monte Carlo (MCMC) technique 
to draw random parameter sets from the probability distributions constructed from the measured data: 
i.e., it returns posterior parameter distributions. 
In this work, we regard the posterior mode and the range 
around it that encompasses the 68\% fraction of the distribution 
as our best-fit value and uncertainty for each regression parameter. 
We used $3 \times 10^4$ random draws returned by the MCMC sampler 
with the Metropolis-Hastings algorithm. 
Note that the \verb|linmix_err| models the prior distribution of 
the independent variable using a weighted mixture of $K$-Gaussians. 
Although \citet{2007ApJ...665.1489K} recommended using $K = 3$ 
to be flexible enough for a wide variety of distributions, 
we instead used $K = 1$, considering the small numbers of samples. 
The canonical Spearman rank correlation coefficient and the corresponding 
null-hypothesis probability were also derived with \verb|r_correlate| in IDL 
to simplify the results.

\section{Results from regression analysis}\label{sec4}
\subsection{Positive $M_{\rm dense}-\dot{M}_{\rm BH}$ correlation}\label{sec4.1}
\begin{figure*}
\epsscale{1}
\plotone{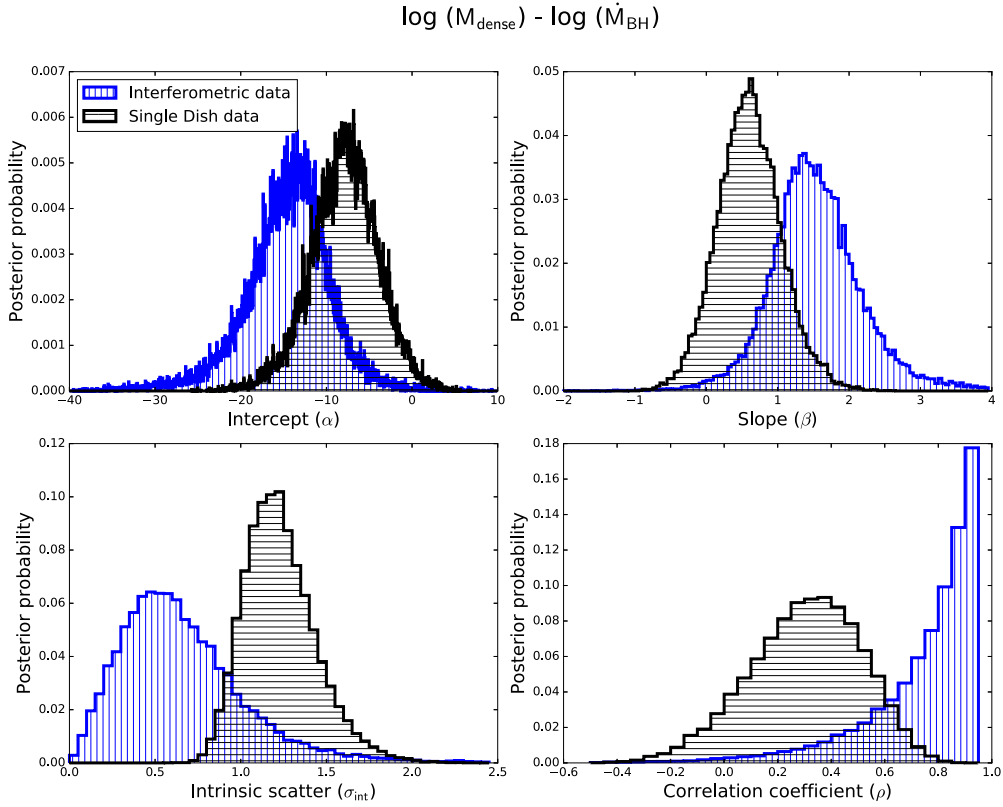}
\caption{
Posterior distributions for the intercept ($\alpha$), slope ($\beta$), intrinsic scatter ($\sigma_{\rm int}$), and correlation coefficient ($\rho$) 
for the $\log (M_{\rm dense})-\log (\dot{M}_{\rm BH})$ relationship (Figure \ref{figure3}b). 
Colors indicate the IT (blue, vertical line) and SD (black, horizontal line) samples, respectively. 
The width of each bin is fixed to 0.05 in these plots. 
Note that the regression analysis for the $\log (L'_{\rm HCN})-\log (L_{\rm 2-10keV})$ yields essentially similar parameter distributions. 
}
\label{figure4}
\end{figure*}

Based on the method described in Section \ref{sec3}, 
we achieved the posterior distribution of each regression parameter defined in Equation (\ref{eq4}). 
For example, Figure \ref{figure4} shows the distributions for the $M_{\rm dense}-\dot{M}_{\rm BH}$ correlation. 
The resulting best-fit parameters are summarized in Table \ref{tbl3} 
and the regression lines constructed with these values are overlaid on Figure \ref{figure3} as the best-fit lines. 
We now focus on the $M_{\rm dense}-\dot{M}_{\rm BH}$ correlation 
because it gives more physically meaningful information 
than the $L'_{\rm HCN}-L_{\rm 2-10keV}$ correlation, 
although both will yield the same conclusions. 
We confirmed a similar correlation in the flux-flux plane as well. 

The estimated parameters for both the IT and SD samples have relatively 
wide ranges, reflecting their limited statistics. 
Nevertheless, there was a positive correlation between these two quantities, 
particularly for the IT sample, as supported by the positive slope and the high correlation coefficient ($>$0.77); 
that is, {\it{the more gas, the more active the AGN}}. 
Here, the canonical null-hypothesis probability returned by \verb|r_correlate| (0.033) 
also showed statistical significance at the 5\% level. 
Because $\dot{M}_{\rm BH}$ is seemingly independent of $M_{\rm BH}$ (\citealt{2004A&A...426..797C}), 
our results suggest the importance of CNDs as the external drivers of AGN activity. 

\subsection{Virtual equivalence to the ${\rm SFR}-\dot{M}_{\rm BH}$ correlations}\label{sec4.2}
To ignite and maintain AGN activity, a sufficient amount of molecular gas (= fuel) 
and the presence of physical mechanisms to cause mass accretion are required. 
The former condition is surely satisfied for our sample 
because the measured $M_{\rm dense}$ of the CND is enough 
to keep the current $\dot{M}_{\rm BH}$ well beyond 100 Myr. 
Given the low $\dot{M}_{\rm BH}$ ($\la 0.1$ $M_\odot$ yr$^{-1}$) 
and the likely prevalence of dusty, compact structures 
in the central regions of Seyfert galaxies (\citealt{2003ApJ...589..774M,2007ApJ...655..718S}), 
indicating the coexistence of gaseous CNDs, 
as observed in our sample galaxies, 
then the first condition could be satisfied in most Seyfert galaxies as well. 

However, measured circumnuclear (i.e., $\la 100$ pc scale) SFRs 
tend to far exceed $\dot{M}_{\rm BH}$, 
by more than $\sim 10$ times, in Seyfert galaxies (e.g., \citealt{2012ApJ...746..168D}). 
This indicates that most of the dense gas in the CNDs 
would be consumed by such star formation. 
Even so, $M_{\rm dense}$ is massive enough 
to keep the current nuclear activity (i.e., $\dot{M}_{\rm BH} + {\rm SFR}$) 
over $\sim 10-100$ Myr so long as there is no massive outflow. 
From this perspective, we can state that the $M_{\rm dense}-\dot{M}_{\rm BH}$ correlation in Figure \ref{figure3} 
is virtually equivalent to the ${\rm SFR}-\dot{M}_{\rm BH}$ correlations 
(\citealt{2012ApJ...746..168D,2014ApJ...780...86E}). 
Thus, our results support the existence of 
the ${\rm SFR}-\dot{M}_{\rm BH}$ correlation in an individual way. 
One advantage of using cold molecular measurements is that, 
today, we can achieve quite high resolutions with ALMA. 
This is essential in probing the CND scale of relatively distant galaxies. 
Moreover, less spectral contamination from an AGN itself 
is expected at millimeter/submillimeter bands than at optical/IR ones. 
Note that we can reproduce the slope of the ${\rm SFR}-\dot{M}_{\rm BH}$ correlations at the CND scale 
using that of the $M_{\rm dense}-\dot{M}_{\rm BH}$ correlation 
and the linear conversion from $M_{\rm dense}$ to SFR (\citealt{2004ApJ...606..271G}). 
We confirmed that this was not the case for the CO-derived molecular gas mass, 
which is also convertible to SFR (\citealt{2012ARA&A..50..531K}). 
This supports the more appropriate use of the HCN($1-0$) emission line 
to rebuild the ${\rm SFR}-\dot{M}_{\rm BH}$ correlation, as described in Section \ref{sec1.1}. 

Our regression analysis also revealed that the correlation was tighter 
for the IT sample than the SD sample in a statistical sense (Figure \ref{figure3}, Table \ref{tbl3}). 
That is, the IT sample showed a smaller scatter ($\sigma_{\rm int} = 0.52^{+0.34}_{-0.29}$ dex) 
and a higher correlation coefficient ($\rho >0.77$) 
than the SD sample ($\sigma_{\rm int} = 1.19^{+0.18}_{-0.21}$ and $\rho = 0.35^{+0.18}_{-0.24}$), 
in the $M_{\rm dense}-\dot{M}_{\rm BH}$ plane. 
This dependence again traces that of the ${\rm SFR}-\dot{M}_{\rm BH}$ correlations 
and indicates that CND-scale gas can be an external regulator of mass accretion further inward, 
whereas an entire galactic-scale does not. 

\begingroup
\renewcommand{\arraystretch}{1.5}
\begin{table*}
\begin{center}
\caption{Results from the regression analysis}\label{tbl3}
\begin{tabular}{cccccccc}
\tableline\tableline
Data & Median aperture & $N_{\rm sample}$ & $\alpha$ & $\beta$ & $\sigma_{\rm int}$ & $\rho$ & $p$ \\
(1) & (2) & (3) & (4) & (5) & (6) & (7) & (8) \\ \hline \hline
\multicolumn{8}{c}{$\log (L'_{\rm HCN})-\log (L_{\rm 2-10keV})$ correlation} \\  \hline
IT-sample & 220 pc & 9 & $34.77^{+2.64}_{-2.63}$ & $1.06^{+0.40}_{-0.39}$ & $0.59^{+0.30}_{-0.23}$ & $0.89^{+0.07}_{-0.23}$ & 0.058 \\
SD-sample & 5.5 kpc & 28 & $38.29^{+2.52}_{-2.52}$ & $0.46^{+0.33}_{-0.32}$ & $1.04^{+0.19}_{-0.14}$ & $0.31^{+0.19}_{-0.19}$ & 0.210 \\ \hline \hline
\multicolumn{8}{c}{$\log (M_{\rm dense})-\log (\dot{M}_{\rm BH})$ correlation} \\  \hline
IT-sample & 220 pc & 10 & $-13.45^{+3.90}_{-5.25}$ & $1.41^{+0.70}_{-0.56}$ & $0.52^{+0.34}_{-0.29}$ & $>$0.77 & 0.033 \\
SD-sample & 5.5 kpc & 29 & $-7.80^{+3.77}_{-3.76}$ & $0.62^{+0.42}_{-0.52}$ & $1.19^{+0.18}_{-0.21}$ & $0.35^{+0.18}_{-0.24}$ & 0.270 \\ \hline 
\tableline
\end{tabular}
\tablecomments{We assumed the formulation of $\log \zeta_i = \alpha + \beta \times \log \xi_i + \epsilon_i$ (Section \ref{sec3}) 
for two regression pairs of ($\xi$, $\zeta$) = ($L'_{\rm HCN}$, $L_{\rm 2-10keV}$) and ($M_{\rm dense}$, $\dot{M}_{\rm BH}$). 
Column 1: type of the sample. 
Column 2: the median aperture for HCN($1-0$) measurement. 
Column 3: number of the sample used for the analysis. Data with an upper limit on $L'_{\rm HCN}$ (or $M_{\rm dense}$) are excluded. 
Columns $4-7$: the intercept ($\alpha$), slope ($\beta$), intrinsic scatter ($\sigma_{\rm int}$), and correlation coefficient ($\rho$) of the regression. 
The quoted values correspond to the mode of the posterior distribution and the range around it that encompasses 68\% fraction of the distribution, 
based on the estimation from the IDL code linmix\_err (\citealt{2007ApJ...665.1489K}). 
Note that the distribution of $\rho$ of the IT-sample does not show a clear turnover 
for the case of $\log (M_{\rm dense})-\log (\dot{M}_{\rm BH})$ correlation. 
We thus show the lower limit of $\rho$ instead, i.e., $0.77 \leq \rho \leq 1.00$ contains 68\% of the posterior distribution. 
Column 8: null-hypothesis probability returned by the IDL code r\_correlate.} 
\end{center}
\end{table*}
\endgroup

\section{Discussion}\label{sec5}
In the previous section, we showed that there is a positive and fairy strong 
correlation between $M_{\rm dense}$ and $\dot{M}_{\rm BH}$ at CNDs, 
indicating that a more active AGN resides in a more gas-rich CND. 
However, as is also the case for the ${\rm SFR}-\dot{M}_{\rm BH}$ correlation, 
this result does not indicate the physical mechanism(s) of 
mass accretion at this spatial scale. 
Thus, here we try to speculate on accretion mechanisms (or the triggers of AGN activity) 
particularly from the perspective of the {\it{AGN$-$starburst connection}}. 
This could lead to important progress in the study of AGN fueling 
because accretion processes (even the dominant one) at $\la 100$ pc scale remain unknown. 
Note that simple Eddington-limited accretion does not explain these Seyfert galaxies 
because the observed $\lambda_{\rm Edd}$ are well below unity (Figure \ref{figure1}). 

To date, one of the most compelling pieces of evidence for a {\it{causal}} connection 
between AGN and (circumnuclear) starburst activities 
would be the observed $\sim 50-100$ Myr time delay between the onset of each activity 
(\citealt{2007ApJ...671.1388D,2009ApJ...692L..19S,2010MNRAS.405..933W}). 
This supports the view that the preceding star formation event subsequently provides fuel for the AGN. 
With this in mind, we focused on the following scenario 
showing a {\it{direct}} connection between AGN and starburst: 
{\it{angular momentum loss due to SN-driven turbulence}} 
(\citealt{2002ApJ...566L..21W,2008ApJ...681...73K,2009ApJ...702...63W}). 
Note that there are other scenarios that also support a direct AGN$-$starburst connection, 
such as {\it{mass loss from evolved stars}} (\citealt{1988ApJ...332..124N,1991ApJ...376..380C,1993ApJ...416...26P,2007ApJ...671.1388D}). 
Consequently, our discussion does not provide a comprehensive, complete view of CND-scale accretion.   
Rather, we seek to demonstrate one possible approach to tackle 
the fueling problem, which may be expanded much greater detail 
with future high-resolution, high-sensitivity observations provided by ALMA. 

\subsection{Angular momentum loss due to SN-driven turbulence}\label{sec5.2}
Type II SNe occur after a certain time delay from the onset of the starburst episode. 
That delay is $10-50$ Myr for a starburst event 
with a $e$-folding time scale ($\tau_{\rm SB}$) of 10 Myr (\citealt{2007ApJ...671.1388D}). 
The delay naturally becomes longer as $\tau_{\rm SB}$ increases. 
Indeed, along with large uncertainty, $\tau_{\rm SB}$ can be as long as 
$\ga 100$ Myr (e.g., \citealt{2009ApJ...692L..19S,2010MNRAS.405..933W,2013ApJ...765...78A}). 
Thus, SN-driven accretion may explain the observed $\sim 50-100$ Myr delay 
between the onsets of starbursts and AGNs. 

According to the numerical simulations of \citet{2002ApJ...566L..21W} and \citet{2009ApJ...702...63W}, 
SNe inject strong turbulence into a CND, 
which effectively removes the angular momentum of the gas and increases $\dot{M}_{\rm BH}$. 
From the observational side, a circumnuclear SFR is typically $0.1-1$ $M_\odot$ yr$^{-1}$ 
(\citealt{2012ApJ...746..168D}) for the IT sample. 
Integrating the Salpeter initial mass function over 
a stellar mass range of $0.1-125$ $M_\odot$ yields 
a type II SN rate of $\sim 0.007$ SFR (\citealt{2009ApJ...702...63W}). 
Then, the current SN rate of the IT sample is $\sim 10^{-3}-10^{-2}$ yr$^{-1}$. 
Because even a CND with a low SN rate ($5.4 \times 10^{-5}$ yr$^{-1}$) 
shows a highly turbulent motion in the numerical simulation (\citealt{2009ApJ...702...63W}), 
it would seem reasonable to expect turbulence-driven 
accretion to occur in the sample CNDs. 
Moreover, kinematic analysis of the H$_2$ $1-0$ S(1) emission line clearly revealed 
that CNDs are highly turbulent,  with a velocity dispersion 
of $\ga 50$ km s$^{-1}$ (\citealt{2009ApJ...696..448H}). 

Then we applied an analytical form of the SN-driven accretion model 
developed by \citet{2008ApJ...681...73K} to our observations. 
A typical ${\rm SFR}/\dot{M}_{\rm BH}$ is a few$-10$ in this model (see also \citealt{2009ApJ...706..676K}), 
which is consistent with the observations. 
Based on the viscous accretion model (\citealt{1981ARA&A..19..137P}), 
they derived a mass accretion rate expected 
at the innermost radius ($r_{\rm in}$) of a CND as
\begin{equation}\label{eq5}
\left( \frac{\dot{M}_{\rm acc}(r_{\rm in})}{\rm{M_\odot~ yr^{-1}}} \right) = 3 \pi \alpha_{\rm SN} \mu_{\rm SN} E_{\rm SN} C_* \Sigma_{\rm dense} (r_{\rm in})\left( \frac{r^3_{\rm in}}{GM_{\rm BH}} \right) .
\end{equation}
In this model, the turbulent pressure is assumed to be in 
hydrodynamical equilibrium with gravity in the vertical direction. 
The so-called viscous-alpha parameter was invoked to express the viscous coefficient 
as $\nu_{\rm t}(r) = \alpha_{\rm SN} \upsilon_{\rm t}(r) h(r)$, 
where $\upsilon_{\rm t}(r)$ and $h(r)$ are the turbulent velocity 
and the scale height of the disk at radius $r$ from the center. 
$E_{\rm SN}$ is the total energy injected by a single SN (i.e., 10$^{51}$ erg), 
and $\mu_{\rm SN}$ is the heating efficiency per unit mass 
that bridges $E_{\rm SN}$ and the kinetic energy of the matter. 
This energy input from SNe is balanced with dissipation due to the turbulence. 
$C_*$, $\Sigma_{\rm dense}$, and $G$ are the star-formation efficiency ($\equiv {\rm SFR}/M_{\rm dense}$), 
the gas surface-density of the disk\footnote{We use the subscript {\it{dense}} considering the high densities of the CNDs.}, 
and the gravitational constant, respectively. 

Regarding the dependence of Equation (\ref{eq5}) on each parameter, 
we emphasize the following two points that match observations well. 
\begin{itemize}
\item {\it{H$_2$ surface brightness:}} Recent VLT integral field unit observations of 
the 2.12 $\mu$m H$_2$ $1-0$ S(1) emission line revealed that 
CNDs of active galaxies show systematically higher H$_2$ surface brightness 
than those of inactive galaxies (\citealt{2013ApJ...768..107H}). 
If we assume enhanced H$_2$ is a reflection of gas mass, 
this is consistent with the form of $\dot{M}_{\rm acc} \propto \Sigma_{\rm dense}$. 
\item {\it{Black hole mass:}} From Tables \ref{tbl1} and \ref{tbl2}, we found NGC 1097 and NGC 2273 
have comparable $M_{\rm dense}$, whereas their $\dot{M}_{\rm BH}$ are totally different, i.e., 
NGC 1097 shows a $\sim 150$ times smaller value. 
However, NGC 1097 has a $\sim 20$ times larger $M_{\rm BH}$ than NGC 2273. 
Thus, we expect that including $M_{\rm BH}$ in a negative form into $\dot{M}_{\rm BH}$ like $\propto M^{-1}_{\rm BH}$ 
would better reproduce the trend in $\dot{M}_{\rm BH}$. 
This is indeed the proposed formulation of $\dot{M}_{\rm acc}$. 
\end{itemize}
These points provided our motivation to 
apply Equation (\ref{eq5}) to the actual observations. 
The negative dependence on $M_{\rm BH}$ indicates that a suppressed scale height of the disk, 
due to strong gravity from the SMBH itself, reduced the turbulent viscosity 
inside the disk, and $\dot{M}_{\rm acc}$ accordingly. 
We also note that Equation (\ref{eq5}) does not necessarily 
disagree with the results of \citet{2004A&A...426..797C}, 
who claimed that black hole mass accretion seemed to be independent of $M_{\rm BH}$, 
because we now include not only $M_{\rm BH}$ but also $\Sigma_{\rm dense}$ as controlling parameters of mass accretion. 

One might question whether ongoing star formation can have 
a direct link with the ongoing AGN activity in terms of the time scale. 
According to \citet{2008ApJ...681...73K}, the viscous time scale 
from the radius $r$ of the CND ($\tau_{\rm vis}$) can be determined by the viscous coefficient $\nu_{\rm t}$ as 
\begin{equation}\label{eq6}
\tau_{\rm vis} = \frac{r^2}{\nu_{\rm t}} = \frac{r^2}{\alpha_{\rm SN} \upsilon_{\rm t}(r) h(r)}
\end{equation}
Putting $r = 30$ pc, $\upsilon_{\rm t}$($r = 30$ pc) = 50 km s$^{-1}$, $h$($r = 30$ pc) = 30 pc 
(these are typically observed values in H$_2$ $1-0$ S(1) disks; 
\citealt{2009ApJ...696..448H}) and $\alpha_{\rm SN} = 1$, 
we obtain $\tau_{\rm vis} \sim 1$ Myr, 
which is comparable to the dynamical time. 
However, using the free-fall time from the radius $r'$ to the center ($\tau_{\rm ff}$), 
the viscous time scale inside the accretion disk is 
\begin{equation}\label{eq7}
\tau_{\rm vis} \sim \frac{\tau_{\rm ff}}{\alpha_{\rm disk} (h/r')^2} ,
\end{equation}
where $\alpha_{\rm disk}$ denotes the viscous alpha-parameter 
in the accretion disk (\citealt{1973A&A....24..337S,1981ARA&A..19..137P}). 
We consider the accretion disk with $r' = 0.1$ pc 
(typical size of water maser disks; \citealt{2011ApJ...727...20K}) 
around the SMBH with $10^7$ $M_\odot$. 
Recent numerical simulations reported $\alpha_{\rm disk} = 0.01-0.1$ 
due to magneto-rotational instability in the disk (e.g., \citealt{2003ApJ...585..429M}). 
Thus, putting $\alpha_{\rm disk} = 0.1$ and $(h/r') = 0.01$ 
(typical value assumed in standard disk models) in Equation (\ref{eq7}), 
we obtain $\tau_{\rm vis} \sim 20$ Myr. 
Consequently, we suggest that the total accretion time scale 
from the CND to the center is at most a few $\times 10$ Myr. 
This is smaller than the typical $\tau_{\rm SF}$ ($\ga 100$ Myr, 
\citealt{2009ApJ...692L..19S,2010MNRAS.405..933W,2013ApJ...765...78A}). 
Therefore, it does seem to be plausible for CNDs and the star formation that occurs inside them 
to have a {\it{causal}} link with on-going AGN activities.

\subsubsection{Comparison with observations}\label{sec.5.2.1}
Next we compared the model-predicted accretion rates with the observed values. 
We restricted the analysis to the IT sample only, 
because galactic-scale molecular gas would have no relevance 
to current AGN activity, as we saw in Section \ref{sec4}. 
Again, we note that this discussion is rather speculative with large uncertainties, 
and should be tested with larger samples based on future observations. 
One important issue is that most of the CNDs are not resolved spatially at the HCN($1-0$) emission line, 
so we cannot measure $\Sigma_{\rm dense}$ directly. 
Thus, we simply assume that the dense gas 
is uniformly distributed within a CND and then reduce Equation (\ref{eq5}) to 
\begin{equation}\label{eq8}
\begin{split}
\left( \frac{\dot{M}_{\rm acc}(r_{\rm in})}{M_\odot ~{\rm yr^{-1}}} \right) &\sim 0.13 \left( \frac{\alpha_{\rm SN}}{1} \right) 
\left( \frac{r_{\rm in}}{\rm{3 ~pc}} \right)^3 \left( \frac{r_{\rm out}}{\rm{30 ~pc}} \right)^{-2} \\ 
&\quad \cdot \left( \frac{C_*}{\rm{10^{-7} ~yr^{-1}}} \right) \left( \frac{M_{\rm dense}}{M_{\rm BH}} \right) .
\end{split}
\end{equation}
Here, the model-simulated fiducial values of $\mu_{\rm SN} = 10^{-3}$ $M^{-1}_\odot$ 
and $\alpha_{\rm SN} \sim 1$ are used (\citealt{2002ApJ...566L..21W}). 

Regarding the outer radii ($r_{\rm out}$) of disks, 
we assume that they are identical to the HWHM of the H$_2$ $1-0$ S(1) 
emission disks (\citealt{2009ApJ...696..448H}). 
The values are listed in Table \ref{tbl4} after correcting the distance to each object. 
A basis of this assumption is the fact that a 100 pc scale spatial distributions of 
the H$_2$ $1-0$ S(1) emission line and those of 
the CO and HCN emission lines are almost identical 
in NGC 1068 (\citealt{2009ApJ...691..749M,2014A&A...567A.125G}) 
and in NGC 1097 (\citealt{2009ApJ...696..448H}; Kohno et al. in prep.). 
With this constraint, we will focus on the IT sample 
with the high-resolution H$_2$ $1-0$ S(1) kinematic information in \citet{2009ApJ...696..448H}, 
specifically, NGC 1097, NGC 3227, NGC 4051, and NGC 7469, in the following analysis. 
Note that we excluded NGC 1068 from this specific investigation because 
the innermost regions of its CND show considerably complex substructures 
that are likely reflecting strong AGN feedback events (\citealt{2009ApJ...696..448H}), 
which violates our assumption of the uniform gas distribution inside a CND. 

For the disk inner radius ($r_{\rm in}$), \citet{2008ApJ...681...73K} simply defined it 
as the dust sublimation radius, 
which we believe has limited meaning. 
Because we are now trying to apply the {\it{SN-driven}} turbulence model, star formation must occur prior to SN explosions. 
However, we would not expect star formation in the gas disk with a temperature as high as, for example, $> 1000$ K. 
Alternatively, we consider two other radii. 
The first is the radius at which the fractional abundance ($f$) 
of H$_2$ becomes equivalent to that of HI, i.e., $f$(H$_2$) = $f$(HI). 
We argue this because stars are born from molecular gas. 
In close vicinity to AGNs, strong X-ray irradiation can substantially alter gas physics and chemistry 
(the X-ray dominated region = XDR; \citealt{1996A&A...306L..21L,1996ApJ...466..561M}). 
Thus, we calculate this radius (denoted as $r_{\rm X}$) following Equation (2) of \citet{1996ApJ...466..561M}. 
According to their calculation, we can estimate $r_{\rm X}$ as the radius 
at which the ratio of X-ray energy deposition rate per particle takes $\log(H_{\rm X}/n_{\rm H}) = -27.5$. 
Note that in the region with $\log(H_{\rm X}/n_{\rm H}) < -27.5$, the gas temperature is $\sim 10s-100$ K. 
A gas volume density of $n_{\rm H_2} = 10^6$ cm$^{-3}$ (here, we regard $n_{\rm H} = n_{\rm H_2}$) 
is adopted based on a recent high-resolution 
multi-transitional study towards the center of NGC 1068 (\citealt{2014A&A...570A..28V}), 
although another value in the range of $n_{\rm H_2} \sim 10^{5-7}$ cm$^{-3}$ would not change our argument. 
The X-ray luminosity in Table \ref{tbl1} was used here to derive $r_{\rm X}$, listed in Table \ref{tbl4}. 
However, even if molecular hydrogen exists, 
stars cannot be formed unless the gas becomes gravitationally unstable. 
Thus, we estimated the radius at which the Toomre-Q parameter (e.g., \citealt{1964ApJ...139.1217T}) became unity, i.e., 
\begin{equation}\label{eq9}
Q \equiv \frac{\kappa(r) c_s}{\pi G \Sigma_{\rm dense}} = 1, 
\end{equation}
where $\kappa$ is the epicyclic frequency, $c_s$ ($\sim 1$ km s$^{-1}$) 
is the sound speed, and $G$ is the gravitational constant. 
The spatially averaged gas surface density $\Sigma_{\rm dense} \sim M_{\rm dense}/\pi r^2_{\rm out}$ was used. 
The radius $r_{\rm Q}$ in Table \ref{tbl4} was derived in this way. 
Using these two radii, we regard the {\it{effective}} innermost 
radius as $r_{\rm in,eff} = {\rm max}[r_{\rm X}, r_{\rm Q}]$. 

To obtain $C_*$, we used SFR measured with a comparable aperture 
to that of the HCN($1-0$) measurements, as $C_* = {\rm SFR}/M_{\rm dense}$. 
This SFR was estimated with the 11.3 $\mu$m PAH emission (\citealt{2012ApJ...746..168D,2014ApJ...780...86E}). 
Because of the mechanism invoked by the model, 
we consider that there is no star formation inside $r_{\rm in,eff}$. 

\begin{figure*}
\epsscale{1}
\plotone{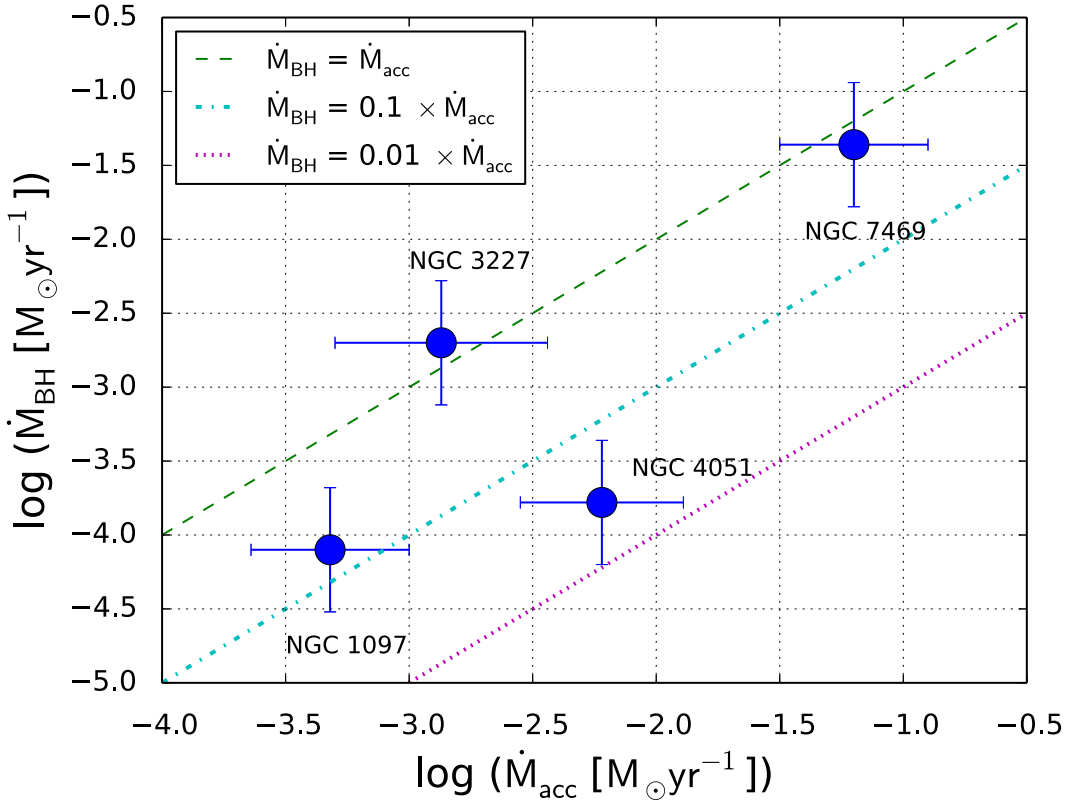}
\caption{
Scatter plot of the mass accretion rate at the {\it{effective}} innermost radius ($r_{\rm in,eff}$) of the CNDs ($\dot{M}_{\rm acc}$) 
predicted by the SN-driven turbulence model (\citealt{2008ApJ...681...73K}), 
and the black hole accretion rate ($\dot{M}_{\rm BH}$). 
The overlaid lines indicate that $\dot{M}_{\rm BH}$ is 100\% (green, dashed), 10\% (cyan, dot-dashed), and 1\% (magenta, dotted) of $\dot{M}_{\rm acc}$. 
In NGC 3227 and NGC 7469, we found $\dot{M}_{\rm BH} \sim \dot{M}_{\rm acc}$, 
indicating that most of the mass accreted from $r_{\rm in,eff}$ 
directly went to the black hole accretion. 
However, $\dot{M}_{\rm BH} \ll \dot{M}_{\rm acc}$ in NGC 1097 and NGC 4051. 
Although not all of the uncertainties are explicitly included in both $\dot{M}_{\rm acc}$ and $\dot{M}_{\rm BH}$, 
this discrepancy would suggest other sources of mass flow, e.g., disk winds (see also Figures \ref{figure6} and \ref{figure7}). 
}
\label{figure5}
\end{figure*}

Following the above, we estimated $\dot{M}_{\rm acc}$ with Equation (\ref{eq8}), 
as tabulated in Table \ref{tbl4}. 
The uncertainty was assumed to be primarily driven by that of $M_{\rm dense}/M_{\rm BH}$ for simplicity, 
because that of the other parameters is not well constrained. 
Note that this $\dot{M}_{\rm acc}$ is now an accretion rate at $r_{\rm in,eff}$. 
Figure \ref{figure5} compares the resulting $\dot{M}_{\rm acc}$ with the observed $\dot{M}_{\rm BH}$. 
According to this SN-driven accretion model, 
one can find $\dot{M}_{\rm acc} \sim \dot{M}_{\rm BH}$ in NGC 7469 and NGC 3227, 
indicating that most of the mass accreted from their $r_{\rm in,eff}$ goes to black hole accretion. 
In contrast to these objects, $\dot{M}_{\rm BH}$ was only $\sim 10\%$ or less of $\dot{M}_{\rm acc}$ in NGC 1097 and NGC 4051, respectively. 
Thus, only a small fraction of the accreting gas from the CND contributed to growing the SMBH in these galaxies. 
In the latter case, we need some other process(es) 
to account for the missing mass flows.

\begin{table*}
\begin{center}
\caption{Parameters of the CNDs for the model prediction \label{tbl4}}
\begin{tabular}{cccccccccc}
\tableline\tableline
Target & $r_{\rm out}$ & $r_{\rm X}$ & $r_{\rm Q}$ & SFR/aperture & Ref. & $C_*$ & $\log(\dot{M}_{\rm acc})$ & $\log(\dot{M}_{\rm wind})$ & Ref. \\
(1) & (2) & (3) & (4) & (5) & (6) & (7) & (8) & (9) & (10) \\ 
\hline \hline
NGC 1097 & 28.8 & 0.38 & 2.1 & 0.14/250 & 1 & 2.7$\times$10$^{-9}$ & -3.32$\pm$0.32 & - & - \\
NGC 3227 & 34.9 & 0.95 & 2.0 & 0.07/60 & 2 & 3.4$\times$10$^{-9}$ & -2.87$\pm$0.43 & - & - \\
NGC 4051 & 7.6 &  0.46 & 0.43 & 0.11/140 & 1 & 2.9$\times$10$^{-8}$ & -2.22$\pm$0.33 & -2.39 & 3,4 \\
NGC 7469 & 30.8 & 2.2 & 0.84 & 1.19/260 & 2 & 6.0$\times$10$^{-9}$ & -1.20$\pm$0.30 & -1.22 & 5 \\
\tableline
\end{tabular}
\tablecomments{
Column 1: name of the sample galaxy that has both the interferometric HCN(1-0) emission data 
and high resolution H$_2$ $1-0$ S(1) data obtained with integral field unit instruments (\citealt{2009ApJ...696..448H}). 
Column 2: the outermost radius of the CND in the unit of pc. 
Column 3: the critical radius that satisfies the condition of $f_{\rm H_2} = f_{\rm HI}$ from XDR calculations (\citealt{1996ApJ...466..561M}). 
Column 4: the critical radius in the unit of pc that satisfies the condition of $Q = 1$. 
Column 5: estimated SFR [$M_\odot$ yr$^{-1}$] based on the 11.3 $\mu$m PAH emission 
and the (geometrically averaged) circular aperture for the measurement in the unit of pc. 
Column 6: reference for Column 5. 
Column 7: estimated SFE in the unit of yr$^{-1}$ by using SFR in Column 5 and $M_{\rm dense}$ of the CND as $C_* = {\rm SFR}/M_{\rm dense}$. 
Column 8: logarithmic scale value of mass accretion rate at $r_{\rm in,eff} = {\rm max}[r_{\rm Q}, r_{\rm X}]$ 
based on the SN-driven turbulence model (\citealt{2008ApJ...681...73K}). 
We assume that the uncertainty is driven by that of $M_{\rm dense}/M_{\rm BH}$ for simplicity. 
Columns 9 and 10: logarithmic scale value of the observed mass outflow rate 
in the X-ray warm absorber and UV absorber found in literature, and the reference for them. 
References: (1) \citet{2012ApJ...746..168D}, (2) \citet{2014ApJ...780...86E}, 
(3) \citet{2007ApJ...659.1022K}, (4) \citet{2012ApJ...751...84K}, (5) \citet{2007A&A...466..107B}. 
}
\end{center}
\end{table*}

\subsubsection{Other components of the mass budget$-$gaseous outflow}\label{sec5.2.2}
\begin{figure*}
\epsscale{1}
\plotone{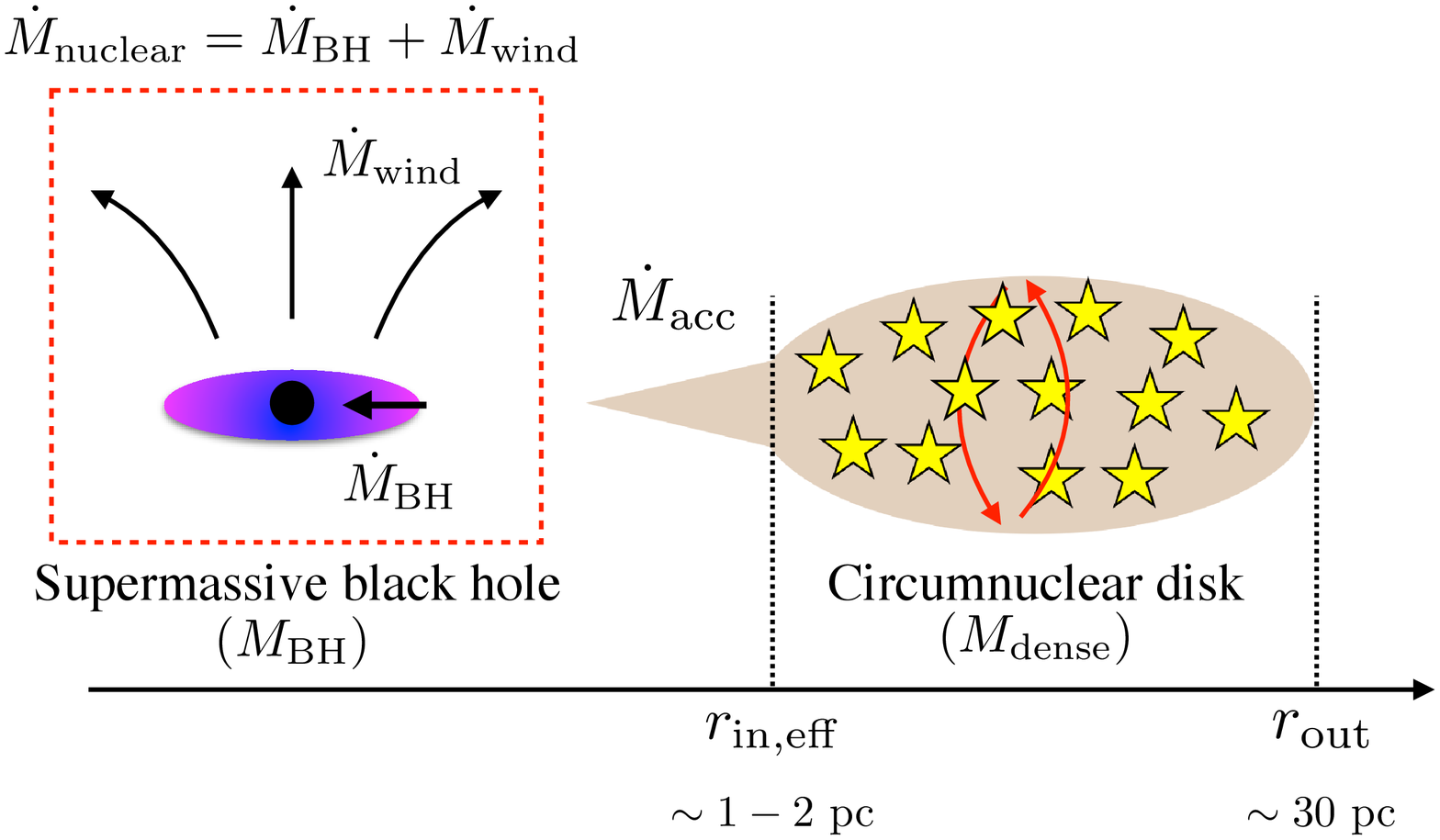}
\caption{
Schematic of the SN-driven turbulent accretion model, 
describing the parameters used for Figures \ref{figure5} and \ref{figure7}. 
The vertical height of the circumnuclear disk (CND) is supported 
by the turbulent pressure due to SN explosions. 
$r_{\rm in,eff}$ and $r_{\rm out}$ are the effective innermost radius 
and the outermost radius, respectively, of the CND (see Section \ref{sec5.2.2} for details). 
This CND, with a gas mass of $M_{\rm dense}$, 
fuels the nuclear (accretion disk-scale) events at the rate of $\dot{M}_{\rm acc}$ 
in the form of either black hole mass accretion ($\dot{M}_{\rm BH}$) or nuclear wind ($\dot{M}_{\rm wind}$). 
We combine these two events into $\dot{M}_{\rm nuclear}$. 
}
\label{figure6}
\end{figure*}

In a more realistic situation, we suggest $\dot{M}_{\rm acc} \neq \dot{M}_{\rm BH}$ in general. 
Indeed, this view is true because the accreted mass from the CND can be expelled in other ways, 
such as nuclear winds (e.g., \citealt{2010RvMP...82.3121G}). 
Such winds may compensate for the discrepancy 
between $\dot{M}_{\rm acc}$ and $\dot{M}_{\rm BH}$ in NGC 1097 and NGC 4051. 
In this scenario, if the mass flow at the accretion disk scale 
can be solely ascribed to either black hole accretion ($\dot{M}_{\rm BH}$) or the winds ($\dot{M}_{\rm wind}$), 
the following relation should hold: 
\begin{equation}\label{eq10}
\dot{M}_{\rm acc} = \dot{M}_{\rm BH} + \dot{M}_{\rm wind} \equiv \dot{M}_{\rm nuclear}. 
\end{equation} 
This situation is described schematically in Figure \ref{figure6}. 

\begin{figure*}
\epsscale{1}
\plotone{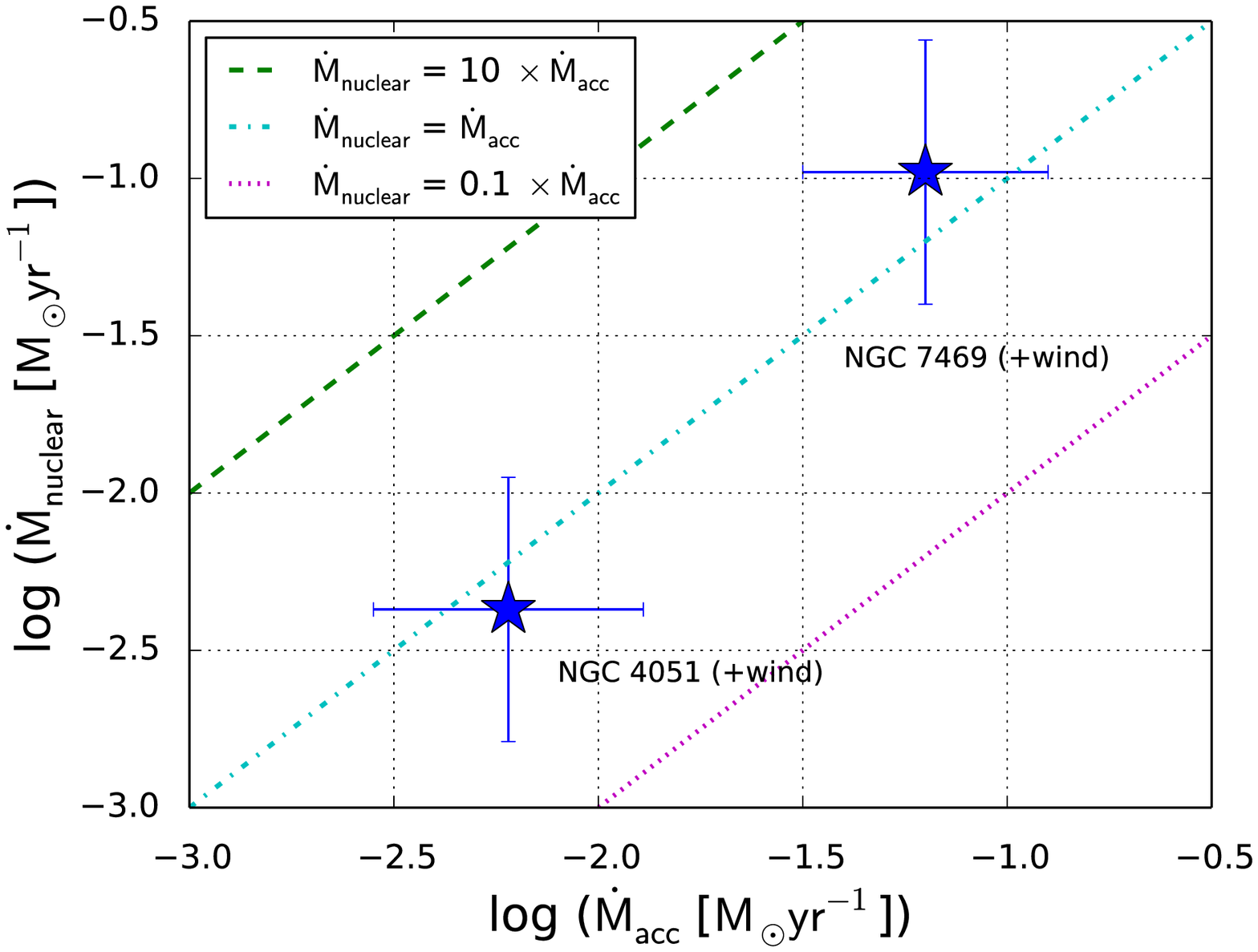}
\caption{
Scatter plot of the mass accretion rate at the {\it{effective}} innermost radius ($r_{\rm in,eff}$) of the CNDs ($\dot{M}_{\rm acc}$) 
predicted by SN-driven turbulence model (\citealt{2008ApJ...681...73K}), 
and the total nuclear mass flow rate at the nuclear (accretion disk-scale) region 
($\dot{M}_{\rm nuclear} \equiv \dot{M}_{\rm BH}$ + $\dot{M}_{\rm wind}$; see also Figure \ref{figure6}). 
The overlaid lines indicate that $\dot{M}_{\rm nuclear}$ is 1000\% (green, dashed), 100\% (cyan, dot-dashed), and 10\% (magenta, dotted) of $\dot{M}_{\rm acc}$. 
There is remarkable agreement between $\dot{M}_{\rm nuclear}$ and $\dot{M}_{\rm acc}$ in NGC 4051 and NGC 7469. 
}
\label{figure7}
\end{figure*}

Unfortunately, $\dot{M}_{\rm wind}$ is available only for NGC 7469 and NGC 4051 (Table \ref{tbl4}) in the literature. 
Here, we considered the winds observed as X-ray/UV warm absorbers 
because these components are highly likely to be located in close vicinity to the accretion disks. 
However, AGN winds observed in atomic or molecular lines 
would be emanating from regions far outside the broad-line region (i.e., the narrow-line region), 
of which $\dot{M}_{\rm wind}$ are substantially mass-loaded by the surrounding ISM. 
Finally, we compare $\dot{M}_{\rm acc}$ with $\dot{M}_{\rm nuclear}$ in Figure \ref{figure7}. 
Only the uncertainty of $\dot{M}_{\rm BH}$ is shown here 
because that of $\dot{M}_{\rm wind}$ is not clearly stated in the references. 
However, we note that the uncertainty of $\dot{M}_{\rm wind}$ 
may be substantially large because it critically depends on the currently unconstrained 
nuclear geometry and volume filling factor of the outflowing gas. 
That aside, it is remarkable that both NGC 7469 and NGC 4051 show 
good agreement of $\dot{M}_{\rm CND} \sim \dot{M}_{\rm nuclear}$ (Figure \ref{figure7}). 
Note that NGC 7469 already showed $\dot{M}_{\rm CND} \sim \dot{M}_{\rm BH}$, 
i.e., $\dot{M}_{\rm wind}$ is not so prominent compared to $\dot{M}_{\rm BH}$ for this object. 

Based on these results, it seems that we are likely describing 
the balance of mass flows at the nuclear regions of Seyfert galaxies (Equation (\ref{eq10})). 
However, our results are tentative due to the small sample size, 
and are based on simplified assumptions. 
Moreover, not all of the expected uncertainties are included in Figures \ref{figure5} and \ref{figure7}. 
Thus, high-resolution observations of dense molecular gas with ALMA to measure $\Sigma_{\rm dense}$ directly, 
and for example, high-sensitivity UV/X-ray spectroscopy to measure $\dot{M}_{\rm wind}$, 
are needed to further improve our understanding 
of circumnuclear mass accretion processes, 
including other mechanisms beyond this SN-driven accretion model.

\subsection{Implications for the evolution of AGNs}\label{sec5.3}
As shown in Equations (\ref{eq5}) and (\ref{eq8}), $\dot{M}_{\rm acc}$ 
in the SN-driven accretion model depends negatively on $M_{\rm BH}$, 
i.e., $\dot{M}_{\rm acc} \propto M^{-1}_{\rm BH}$. 
This is a totally different dependence from those assumed 
in Eddington-limited accretion ($\propto$ $M_{\rm BH}$) 
and Bondi accretion ($\propto M^2_{\rm BH}$). 
Because of this dependence, for a given $M_{\rm dense}$ (or $\Sigma_{\rm dense}$), 
it becomes more difficult for an SMBH to grow as its $M_{\rm BH}$ becomes larger. 
Observationally, the growth factor of $M_{\rm BH}$ in the coming $\sim 100$ Myr is at most $\sim 2$, 
considering the low $\dot{M}_{\rm BH}$ currently observed, 
unless massive inflows occur from the host galaxies into the CNDs to forcibly enhance $\Sigma_{\rm dense}$. 
Because the total lifetime of AGNs would be $\sim 10^{8-9}$ yr (e.g., \citealt{2004MNRAS.351..169M}), 
most of the SMBH accretion would have occurred in the distant past, 
at the early epoch of their evolution, when $M_{\rm BH} \ll M_{\rm dense}$ was surely satisfied. 
\citet{2004A&A...420L..23K} claimed that most growth of SMBHs occurred in the super-Eddington phase when they were young, 
whereas the growth rate in the sub-Eddington phase is modest. 
Such super-Eddington accretion can be allowed in this SN-driven model, 
for example, for an SMBH with $M_{\rm BH} = 10^{5-6}$ $M_\odot$ 
embedded in a CND of $M_{\rm dense} \ga 10^8$ $M_\odot$, 
judging from Equation (\ref{eq8}). 

However, it is highly challenging 
to form SMBHs with $M_{\rm BH} \ga 10^9$ $M_\odot$ 
typically found in local elliptical galaxies (e.g., \citealt{2013ARA&A..51..511K}) 
and high redshift quasars even at $z >6$ (e.g., \citealt{2011Natur.474..616M}) 
with this SN-driven accretion. 
If such black holes accumulated their mass within $10^9$ yr (\citealt{2004MNRAS.351..169M}), 
the average mass accretion rate would be $\sim 1$ $M_\odot$ yr$^{-1}$. 
Even if we assume a quite high $C_*$, such as $10^{-7}$ yr$^{-1}$, 
which is an upper value observed in submillimeter galaxies (\citealt{2005ARA&A..43..677S}), 
$\ga 10^{10}$ $M_\odot$ molecular gas should be accumulated in the CND within 10$^9$ yr 
and should remain there for almost over the entire period of SMBH growth. 
This mass is comparable to the total (i.e., galactic scale) amount of dense molecular gas 
observed in high redshift quasars (e.g., \citealt{2011ApJ...739L..34W}) as well as nearby galaxies (see the case of the SD sample in Figure \ref{figure3}b). 
This, in turn, indicates the difficulty forming high mass-end SMBHs 
because it is highly unlikely for a galaxy to confine all of its molecular gas inside the CND (central $\sim$ 100 pc). 

Regarding lower-mass SMBHs, by assuming $\sim 10\%$ of the total dense molecular gas of the galaxy is in the CND, 
it would be possible to form an SMBH with $\sim 10^8$ $M_\odot$, 
based on the same speculation described above. 
Black holes in this mass range ($M_{\rm BH} \la 10^8$ $M_\odot$) could have been formed 
following a downsizing trend (\citealt{2003ApJ...598..886U,2014ApJ...786..104U}) 
because high-redshift galaxies tend to show higher $C_*$ (\citealt{2005ARA&A..43..677S}), 
which controls $\dot{M}_{\rm acc}$. 
To make high mass-end SMBHs ($M_{\rm BH} \ga 10^9$ $M_\odot$), 
we need some other mechanism(s), such as mergers of black holes 
or long-lasting mass inflows from intragalactic space.

\section{Conclusions}\label{sec6}
Motivated by the ${\rm SFR}-\dot{M}_{\rm BH}$ correlation, 
(e.g., \citealt{2012ApJ...746..168D,2014ApJ...780...86E}), 
we investigated the correlation between the mass of dense molecular gas 
($M_{\rm dense}$) traced by the HCN($1-0$) emission line 
and the black hole mass accretion rate ($\dot{M}_{\rm BH}$) in nearby Seyfert galaxies, 
with a Bayesian-based regression analysis. 
Most of the data were compiled from the literature and/or the ALMA archive. 
Because of the high spatial resolution provided by the interferometers (PdBI, NMA, and ALMA), 
we could probe molecular gas at the CND-scale. 
Our main conclusions are summarized as follows. 

\begin{itemize}
\item There is a positive correlation between $M_{\rm dense}$ and $\dot{M}_{\rm BH}$ at the CND-scale. 
Because dense molecular gas is the site of star formation, 
this correlation is virtually equivalent to the (nuclear) ${\rm SFR}-\dot{M}_{\rm BH}$ correlations found so far. 
Thus, it seems that we succeeded in rebuilding the ${\rm SFR}-\dot{M}_{\rm BH}$ correlations 
individually from the perspective of cold molecular gas observations. 
\item The $M_{\rm dense}-\dot{M}_{\rm BH}$ correlation is significantly tighter for the sample where 
the HCN($1-0$) emission line was measured at the CND scale (IT sample; median aperture = 220 pc) 
than for the sample measured at the galactic scale (SD sample; median aperture = 5.5 kpc). 
This is again consistent with the trend found in the ${\rm SFR}-\dot{M}_{\rm BH}$ correlations, 
suggesting that CND-scale molecular gas plays an important role in fueling AGNs, 
whereas the galactic-scale molecular gas does not. 
\item Assuming that star formation in CNDs directly provides fuel for AGNs, 
we adopted an SN-driven turbulent accretion scenario (\citealt{2008ApJ...681...73K}). 
Although there are still large observational uncertainties in the parameters needed for the model 
(and not all of the uncertainties are explicitly addressed in this study), 
direct comparisons of the model-predicted $\dot{M}_{\rm acc}$ 
(= mass accretion rate at the innermost radius of the CND to further inwards) 
with $\dot{M}_{\rm BH}$ were conducted. 
We found that only a partial fraction (from $\la 10\%$ to $\sim 100\%$) 
of $\dot{M}_{\rm acc}$ was converted into $\dot{M}_{\rm BH}$ in general. 
\item On the other hand, we also found good agreement between $\dot{M}_{\rm nuclear}$ 
= $\dot{M}_{\rm BH} + \dot{M}_{\rm wind}$ (mass outflow rate as nuclear winds) 
and model-predicted $\dot{M}_{\rm acc}$ in NGC 4051 and NGC 7469. 
This result would suggest that we now might be describing the balance of mass flow 
in the nuclear regions of these Seyfert galaxies, 
although this view is based on the tentative and simplified assumptions in this work. 
We suggest that high-resolution observations of dense molecular gas with ALMA to accurately measure 
$\Sigma_{\rm dense}$, as well as, for example, high-sensitivity spectroscopic observations of nuclear winds 
in the UV/X-ray bands are needed to better understand the validity of this SN-driven accretion model. 
\end{itemize}

Because we used the SN-driven accretion model as a demonstration of 
one possible approach for studying CND-scale accretion processes in the ALMA era, 
we should, of course, test other models (e.g., mass-loss from evolved stars) quantitatively 
by increasing the number of high-resolution measurements of CND-scale dense gas.

\acknowledgments
We gratefully appreciate the comments from 
the anonymous referee that significantly improved this article. 
T. I. also appreciate the useful comments from T. Minezaki, R. Makiya, and M. Kokubo 
at the Institute of Astronomy, The University of Tokyo. 
This paper makes use of the following ALMA data: ADS/JAO.ALMA\#2011.0.00108.S, 2012.1.00165.S, and 2012.1.00456.S. 
ALMA is a partnership of ESO (representing its member states), NSF (USA), and NINS (Japan), 
together with NRC (Canada), NSC and ASIAA (Taiwan), and KASI (Republic of Korea), 
in cooperation with the Republic of Chile. 
The Joint ALMA Observatory is operated by ESO, AUI/NRAO and NAOJ. 
This research has made use of the NASA/IPAC Extragalactic Database (NED) 
which is operated by the Jet Propulsion Laboratory, California Institute of Technology, 
under contract with the National Aeronautics and Space Administration. 
We acknowledge the usage of the convenient HyperLeda database as well. 
Part of the results are based on data extracted from the Japanese Virtual Observatory, 
which is operated by the Astronomy Data Center, National Astronomical Observatory of Japan. 
T. I. and K. K. were supported by the ALMA Japan Research Grant of NAOJ Chile Observatory, 
NAOJ-ALMA-0029, 0075, and 0004, respectively. 
K. K. acknowledges support from JSPS KAKENHI Grant Number 25247019. 
N. K. acknowledges the financial support of Grant-in-Aid for Young Scientists (B:25800099).
T. I. is thankful for the fellowship received from 
the Japan Society for the Promotion of Science 
(JSPS KAKENHI Grant Number 14J08410).

\bibliography{Izumi_CND}

\begin{thebibliography}{}
\expandafter\ifx\csname natexlab\endcsname\relax\def\natexlab#1{#1}\fi

\bibitem[{{Abramowicz} \& {Fragile}(2013)}]{2013LRR....16....1A}
{Abramowicz}, M.~A., \& {Fragile}, P.~C. 2013, Living Reviews in Relativity,
  16, 1

\bibitem[{{Alexander} \& {Hickox}(2012)}]{2012NewAR..56...93A}
{Alexander}, D.~M., \& {Hickox}, R.~C. 2012, \nar, 56, 93

\bibitem[{{Alonso-Herrero} {et~al.}(2013){Alonso-Herrero}, {Pereira-Santaella},
  {Rieke}, {Diamond-Stanic}, {Wang}, {Hern{\'a}n-Caballero}, \&
  {Rigopoulou}}]{2013ApJ...765...78A}
{Alonso-Herrero}, A., {Pereira-Santaella}, M., {Rieke}, G.~H., {et~al.} 2013,
  \apj, 765, 78

\bibitem[{{Antonucci}(1993)}]{1993ARA&A..31..473A}
{Antonucci}, R. 1993, \araa, 31, 473

\bibitem[{{Baan} {et~al.}(2008){Baan}, {Henkel}, {Loenen}, {Baudry}, \&
  {Wiklind}}]{2008A&A...477..747B}
{Baan}, W.~A., {Henkel}, C., {Loenen}, A.~F., {Baudry}, A., \& {Wiklind}, T.
  2008, \aap, 477, 747

\bibitem[{{Ballantyne}(2008)}]{2008ApJ...685..787B}
{Ballantyne}, D.~R. 2008, \apj, 685, 787

\bibitem[{{Barnes} \& {Hernquist}(1992)}]{1992ARA&A..30..705B}
{Barnes}, J.~E., \& {Hernquist}, L. 1992, \araa, 30, 705

\bibitem[{{Bentz} \& {Katz}(2015)}]{2015PASP..127...67B}
{Bentz}, M.~C., \& {Katz}, S. 2015, \pasp, 127, 67

\bibitem[{{Blustin} {et~al.}(2007){Blustin}, {Kriss}, {Holczer}, {Behar},
  {Kaastra}, {Page}, {Kaspi}, {Branduardi-Raymont}, \&
  {Steenbrugge}}]{2007A&A...466..107B}
{Blustin}, A.~J., {Kriss}, G.~A., {Holczer}, T., {et~al.} 2007, \aap, 466, 107

\bibitem[{{Brightman} \& {Nandra}(2011)}]{2011MNRAS.413.1206B}
{Brightman}, M., \& {Nandra}, K. 2011, \mnras, 413, 1206

\bibitem[{{Cheung} {et~al.}(2015){Cheung}, {Trump}, {Athanassoula}, {Bamford},
  {Bell}, {Bosma}, {Cardamone}, {Casteels}, {Faber}, {Fang}, {Fortson},
  {Kocevski}, {Koo}, {Laine}, {Lintott}, {Masters}, {Melvin}, {Nichol},
  {Schawinski}, {Simmons}, {Smethurst}, \& {Willett}}]{2015MNRAS.447..506C}
{Cheung}, E., {Trump}, J.~R., {Athanassoula}, E., {et~al.} 2015, \mnras, 447,
  506

\bibitem[{{Cid Fernandes} {et~al.}(2004){Cid Fernandes}, {Gu}, {Melnick},
  {Terlevich}, {Terlevich}, {Kunth}, {Rodrigues Lacerda}, \&
  {Joguet}}]{2004MNRAS.355..273C}
{Cid Fernandes}, R., {Gu}, Q., {Melnick}, J., {et~al.} 2004, \mnras, 355, 273

\bibitem[{{Ciotti} {et~al.}(1991){Ciotti}, {D'Ercole}, {Pellegrini}, \&
  {Renzini}}]{1991ApJ...376..380C}
{Ciotti}, L., {D'Ercole}, A., {Pellegrini}, S., \& {Renzini}, A. 1991, \apj,
  376, 380

\bibitem[{{Cisternas} {et~al.}(2015){Cisternas}, {Sheth}, {Salvato}, {Knapen},
  {Civano}, \& {Santini}}]{2015ApJ...802..137C}
{Cisternas}, M., {Sheth}, K., {Salvato}, M., {et~al.} 2015, \apj, 802, 137

\bibitem[{{Cisternas} {et~al.}(2011){Cisternas}, {Jahnke}, {Inskip},
  {Kartaltepe}, {Koekemoer}, {Lisker}, {Robaina}, {Scodeggio}, {Sheth},
  {Trump}, {Andrae}, {Miyaji}, {Lusso}, {Brusa}, {Capak}, {Cappelluti},
  {Civano}, {Ilbert}, {Impey}, {Leauthaud}, {Lilly}, {Salvato}, {Scoville}, \&
  {Taniguchi}}]{2011ApJ...726...57C}
{Cisternas}, M., {Jahnke}, K., {Inskip}, K.~J., {et~al.} 2011, \apj, 726, 57

\bibitem[{{Collin} \& {Kawaguchi}(2004)}]{2004A&A...426..797C}
{Collin}, S., \& {Kawaguchi}, T. 2004, \aap, 426, 797

\bibitem[{{Combes} {et~al.}(2014){Combes}, {Garc{\'{\i}}a-Burillo}, {Casasola},
  {Hunt}, {Krips}, {Baker}, {Boone}, {Eckart}, {Marquez}, {Neri}, {Schinnerer},
  \& {Tacconi}}]{2014A&A...565A..97C}
{Combes}, F., {Garc{\'{\i}}a-Burillo}, S., {Casasola}, V., {et~al.} 2014, \aap,
  565, A97

\bibitem[{{Curran} {et~al.}(2000){Curran}, {Aalto}, \&
  {Booth}}]{2000A&AS..141..193C}
{Curran}, S.~J., {Aalto}, S., \& {Booth}, R.~S. 2000, \aaps, 141, 193

\bibitem[{{Curran} {et~al.}(2001){Curran}, {Johansson}, {Bergman},
  {Heikkil{\"a}}, \& {Aalto}}]{2001A&A...367..457C}
{Curran}, S.~J., {Johansson}, L.~E.~B., {Bergman}, P., {Heikkil{\"a}}, A., \&
  {Aalto}, S. 2001, \aap, 367, 457

\bibitem[{{Davies} {et~al.}(2012){Davies}, {Mark}, \&
  {Sternberg}}]{2012A&A...537A.133D}
{Davies}, R., {Mark}, D., \& {Sternberg}, A. 2012, \aap, 537, A133

\bibitem[{{Davies} {et~al.}(2007){Davies}, {M{\"u}ller S{\'a}nchez}, {Genzel},
  {Tacconi}, {Hicks}, {Friedrich}, \& {Sternberg}}]{2007ApJ...671.1388D}
{Davies}, R.~I., {M{\"u}ller S{\'a}nchez}, F., {Genzel}, R., {et~al.} 2007,
  \apj, 671, 1388

\bibitem[{{Davies} {et~al.}(2006){Davies}, {Thomas}, {Genzel}, {M{\"u}ller
  S{\'a}nchez}, {Tacconi}, {Sternberg}, {Eisenhauer}, {Abuter}, {Saglia}, \&
  {Bender}}]{2006ApJ...646..754D}
{Davies}, R.~I., {Thomas}, J., {Genzel}, R., {et~al.} 2006, \apj, 646, 754

\bibitem[{{de Rosa} {et~al.}(2012){de Rosa}, {Panessa}, {Bassani}, {Bazzano},
  {Bird}, {Landi}, {Malizia}, {Molina}, \& {Ubertini}}]{2012MNRAS.420.2087D}
{de Rosa}, A., {Panessa}, F., {Bassani}, L., {et~al.} 2012, \mnras, 420, 2087

\bibitem[{{Diamond-Stanic} \& {Rieke}(2012)}]{2012ApJ...746..168D}
{Diamond-Stanic}, A.~M., \& {Rieke}, G.~H. 2012, \apj, 746, 168

\bibitem[{{Diamond-Stanic} {et~al.}(2009){Diamond-Stanic}, {Rieke}, \&
  {Rigby}}]{2009ApJ...698..623D}
{Diamond-Stanic}, A.~M., {Rieke}, G.~H., \& {Rigby}, J.~R. 2009, \apj, 698, 623

\bibitem[{{Esquej} {et~al.}(2014){Esquej}, {Alonso-Herrero},
  {Gonz{\'a}lez-Mart{\'{\i}}n}, {H{\"o}nig}, {Hern{\'a}n-Caballero}, {Roche},
  {Ramos Almeida}, {Mason}, {D{\'{\i}}az-Santos}, {Levenson}, {Aretxaga},
  {Rodr{\'{\i}}guez Espinosa}, \& {Packham}}]{2014ApJ...780...86E}
{Esquej}, P., {Alonso-Herrero}, A., {Gonz{\'a}lez-Mart{\'{\i}}n}, O., {et~al.}
  2014, \apj, 780, 86

\bibitem[{{Ferrarese} \& {Merritt}(2000)}]{2000ApJ...539L...9F}
{Ferrarese}, L., \& {Merritt}, D. 2000, \apjl, 539, L9

\bibitem[{{Gabor} {et~al.}(2009){Gabor}, {Impey}, {Jahnke}, {Simmons}, {Trump},
  {Koekemoer}, {Brusa}, {Cappelluti}, {Schinnerer}, {Smol{\v c}i{\'c}},
  {Salvato}, {Rhodes}, {Mobasher}, {Capak}, {Massey}, {Leauthaud}, \&
  {Scoville}}]{2009ApJ...691..705G}
{Gabor}, J.~M., {Impey}, C.~D., {Jahnke}, K., {et~al.} 2009, \apj, 691, 705

\bibitem[{{Gao} \& {Solomon}(2004{\natexlab{a}})}]{2004ApJS..152...63G}
{Gao}, Y., \& {Solomon}, P.~M. 2004{\natexlab{a}}, \apjs, 152, 63

\bibitem[{{Gao} \& {Solomon}(2004{\natexlab{b}})}]{2004ApJ...606..271G}
---. 2004{\natexlab{b}}, \apj, 606, 271

\bibitem[{{Garc{\'{\i}}a-Burillo} {et~al.}(2008){Garc{\'{\i}}a-Burillo},
  {Combes}, {Usero}, \& {Graci{\'a}-Carpio}}]{2008JPhCS.131a2031G}
{Garc{\'{\i}}a-Burillo}, S., {Combes}, F., {Usero}, A., \& {Graci{\'a}-Carpio},
  J. 2008, Journal of Physics Conference Series, 131, 012031

\bibitem[{{Garc{\'{\i}}a-Burillo} {et~al.}(2014){Garc{\'{\i}}a-Burillo},
  {Combes}, {Usero}, {Aalto}, {Krips}, {Viti}, {Alonso-Herrero}, {Hunt},
  {Schinnerer}, {Baker}, {Boone}, {Casasola}, {Colina}, {Costagliola},
  {Eckart}, {Fuente}, {Henkel}, {Labiano}, {Mart{\'{\i}}n}, {M{\'a}rquez},
  {Muller}, {Planesas}, {Ramos Almeida}, {Spaans}, {Tacconi}, \& {van der
  Werf}}]{2014A&A...567A.125G}
{Garc{\'{\i}}a-Burillo}, S., {Combes}, F., {Usero}, A., {et~al.} 2014, \aap,
  567, A125

\bibitem[{{Garcia-Rissmann} {et~al.}(2005){Garcia-Rissmann}, {Vega}, {Asari},
  {Cid Fernandes}, {Schmitt}, {Gonz{\'a}lez Delgado}, \&
  {Storchi-Bergmann}}]{2005MNRAS.359..765G}
{Garcia-Rissmann}, A., {Vega}, L.~R., {Asari}, N.~V., {et~al.} 2005, \mnras,
  359, 765

\bibitem[{{Genzel} {et~al.}(2010){Genzel}, {Eisenhauer}, \&
  {Gillessen}}]{2010RvMP...82.3121G}
{Genzel}, R., {Eisenhauer}, F., \& {Gillessen}, S. 2010, Reviews of Modern
  Physics, 82, 3121

\bibitem[{{Greenhill} {et~al.}(1997){Greenhill}, {Moran}, \&
  {Herrnstein}}]{1997ApJ...481L..23G}
{Greenhill}, L.~J., {Moran}, J.~M., \& {Herrnstein}, J.~R. 1997, \apjl, 481,
  L23

\bibitem[{{Greenhill} {et~al.}(2003){Greenhill}, {Booth}, {Ellingsen},
  {Herrnstein}, {Jauncey}, {McCulloch}, {Moran}, {Norris}, {Reynolds}, \&
  {Tzioumis}}]{2003ApJ...590..162G}
{Greenhill}, L.~J., {Booth}, R.~S., {Ellingsen}, S.~P., {et~al.} 2003, \apj,
  590, 162

\bibitem[{{Grier} {et~al.}(2013){Grier}, {Martini}, {Watson}, {Peterson},
  {Bentz}, {Dasyra}, {Dietrich}, {Ferrarese}, {Pogge}, \&
  {Zu}}]{2013ApJ...773...90G}
{Grier}, C.~J., {Martini}, P., {Watson}, L.~C., {et~al.} 2013, \apj, 773, 90

\bibitem[{{G{\"u}ltekin} {et~al.}(2009){G{\"u}ltekin}, {Richstone}, {Gebhardt},
  {Lauer}, {Tremaine}, {Aller}, {Bender}, {Dressler}, {Faber}, {Filippenko},
  {Green}, {Ho}, {Kormendy}, {Magorrian}, {Pinkney}, \&
  {Siopis}}]{2009ApJ...698..198G}
{G{\"u}ltekin}, K., {Richstone}, D.~O., {Gebhardt}, K., {et~al.} 2009, \apj,
  698, 198

\bibitem[{{Heckman} {et~al.}(1995){Heckman}, {Krolik}, {Meurer}, {Calzetti},
  {Kinney}, {Koratkar}, {Leitherer}, {Robert}, \&
  {Wilson}}]{1995ApJ...452..549H}
{Heckman}, T., {Krolik}, J., {Meurer}, G., {et~al.} 1995, \apj, 452, 549

\bibitem[{{Hicks} {et~al.}(2013){Hicks}, {Davies}, {Maciejewski}, {Emsellem},
  {Malkan}, {Dumas}, {M{\"u}ller-S{\'a}nchez}, \&
  {Rivers}}]{2013ApJ...768..107H}
{Hicks}, E.~K.~S., {Davies}, R.~I., {Maciejewski}, W., {et~al.} 2013, \apj,
  768, 107

\bibitem[{{Hicks} {et~al.}(2009){Hicks}, {Davies}, {Malkan}, {Genzel},
  {Tacconi}, {M{\"u}ller S{\'a}nchez}, \& {Sternberg}}]{2009ApJ...696..448H}
{Hicks}, E.~K.~S., {Davies}, R.~I., {Malkan}, M.~A., {et~al.} 2009, \apj, 696,
  448

\bibitem[{{Ho}(2008)}]{2008ARA&A..46..475H}
{Ho}, L.~C. 2008, \araa, 46, 475

\bibitem[{{Hobbs} {et~al.}(2011){Hobbs}, {Nayakshin}, {Power}, \&
  {King}}]{2011MNRAS.413.2633H}
{Hobbs}, A., {Nayakshin}, S., {Power}, C., \& {King}, A. 2011, \mnras, 413,
  2633

\bibitem[{{Honma} {et~al.}(1995){Honma}, {Sofue}, \&
  {Arimoto}}]{1995A&A...304....1H}
{Honma}, M., {Sofue}, Y., \& {Arimoto}, N. 1995, \aap, 304, 1

\bibitem[{{Hopkins}(2012)}]{2012MNRAS.420L...8H}
{Hopkins}, P.~F. 2012, \mnras, 420, L8

\bibitem[{{Hopkins} \& {Hernquist}(2006)}]{2006ApJS..166....1H}
{Hopkins}, P.~F., \& {Hernquist}, L. 2006, \apjs, 166, 1

\bibitem[{{Hopkins} \& {Hernquist}(2009)}]{2009ApJ...694..599H}
---. 2009, \apj, 694, 599

\bibitem[{{Hopkins} {et~al.}(2008){Hopkins}, {Hernquist}, {Cox}, \& {Kere{\v
  s}}}]{2008ApJS..175..356H}
{Hopkins}, P.~F., {Hernquist}, L., {Cox}, T.~J., \& {Kere{\v s}}, D. 2008,
  \apjs, 175, 356

\bibitem[{{Hopkins} \& {Quataert}(2010)}]{2010MNRAS.407.1529H}
{Hopkins}, P.~F., \& {Quataert}, E. 2010, \mnras, 407, 1529

\bibitem[{{Hunt} \& {Malkan}(2004)}]{2004ApJ...616..707H}
{Hunt}, L.~K., \& {Malkan}, M.~A. 2004, \apj, 616, 707

\bibitem[{{Imanishi} \& {Wada}(2004)}]{2004ApJ...617..214I}
{Imanishi}, M., \& {Wada}, K. 2004, \apj, 617, 214

\bibitem[{{Izumi} {et~al.}(2013){Izumi}, {Kohno}, {Mart{\'{\i}}n}, {Espada},
  {Harada}, {Matsushita}, {Hsieh}, {Turner}, {Meier}, {Schinnerer}, {Imanishi},
  {Tamura}, {Curran}, {Doi}, {Fathi}, {Krips}, {Lundgren}, {Nakai}, {Nakajima},
  {Regan}, {Sheth}, {Takano}, {Taniguchi}, {Terashima}, {Tosaki}, \&
  {Wiklind}}]{2013PASJ...65..100I}
{Izumi}, T., {Kohno}, K., {Mart{\'{\i}}n}, S., {et~al.} 2013, \pasj, 65, 100

\bibitem[{{Izumi} {et~al.}(2015){Izumi}, {Kohno}, {Aalto}, {Doi}, {Espada},
  {Fathi}, {Harada}, {Hatsukade}, {Hattori}, {Hsieh}, {Ikarashi}, {Imanishi},
  {Iono}, {Ishizuki}, {Krips}, {Mart{\'{\i}}n}, {Matsushita}, {Meier}, {Nagai},
  {Nakai}, {Nakajima}, {Nakanishi}, {Nomura}, {Regan}, {Schinnerer}, {Sheth},
  {Takano}, {Tamura}, {Terashima}, {Tosaki}, {Turner}, {Umehata}, \&
  {Wiklind}}]{2015ApJ...811...39I}
{Izumi}, T., {Kohno}, K., {Aalto}, S., {et~al.} 2015, \apj, 811, 39

\bibitem[{{Izumi} {et~al.}(2016){Izumi}, {Kohno}, {Aalto}, {Espada}, {Fathi},
  {Harada}, {Hatsukade}, {Hsieh}, {Imanishi}, {Krips}, {Mart{\'{\i}}n},
  {Matsushita}, {Meier}, {Nakai}, {Nakanishi}, {Schinnerer}, {Sheth},
  {Terashima}, \& {Turner}}]{2016ApJ...818...42I}
---. 2016, \apj, 818, 42

\bibitem[{{Jiang} {et~al.}(2011){Jiang}, {Wang}, \& {Gu}}]{2011MNRAS.418.1753J}
{Jiang}, X., {Wang}, J., \& {Gu}, Q. 2011, \mnras, 418, 1753

\bibitem[{{Jogee}(2006)}]{2006LNP...693..143J}
{Jogee}, S. 2006, in Lecture Notes in Physics, Berlin Springer Verlag, Vol.
  693, Physics of Active Galactic Nuclei at all Scales, ed. D.~{Alloin}, 143

\bibitem[{{Kamenetzky} {et~al.}(2014){Kamenetzky}, {Rangwala}, {Glenn},
  {Maloney}, \& {Conley}}]{2014ApJ...795..174K}
{Kamenetzky}, J., {Rangwala}, N., {Glenn}, J., {Maloney}, P.~R., \& {Conley},
  A. 2014, \apj, 795, 174

\bibitem[{{Kaviraj}(2014)}]{2014MNRAS.440.2944K}
{Kaviraj}, S. 2014, \mnras, 440, 2944

\bibitem[{{Kawaguchi} {et~al.}(2004){Kawaguchi}, {Aoki}, {Ohta}, \&
  {Collin}}]{2004A&A...420L..23K}
{Kawaguchi}, T., {Aoki}, K., {Ohta}, K., \& {Collin}, S. 2004, \aap, 420, L23

\bibitem[{{Kawakatu} \& {Wada}(2008)}]{2008ApJ...681...73K}
{Kawakatu}, N., \& {Wada}, K. 2008, \apj, 681, 73

\bibitem[{{Kawakatu} \& {Wada}(2009)}]{2009ApJ...706..676K}
---. 2009, \apj, 706, 676

\bibitem[{{Kelly}(2007)}]{2007ApJ...665.1489K}
{Kelly}, B.~C. 2007, \apj, 665, 1489

\bibitem[{{Kennicutt} \& {Evans}(2012)}]{2012ARA&A..50..531K}
{Kennicutt}, R.~C., \& {Evans}, N.~J. 2012, \araa, 50, 531

\bibitem[{{Kl{\"o}ckner} \& {Baan}(2004)}]{2004A&A...419..887K}
{Kl{\"o}ckner}, H.-R., \& {Baan}, W.~A. 2004, \aap, 419, 887

\bibitem[{{Kohno}(2005)}]{2005AIPC..783..203K}
{Kohno}, K. 2005, in American Institute of Physics Conference Series, Vol. 783,
  The Evolution of Starbursts, ed. S.~{H{\"u}ttmeister}, E.~{Manthey},
  D.~{Bomans}, \& K.~{Weis}, 203--208

\bibitem[{{Kohno} {et~al.}(2003){Kohno}, {Ishizuki}, {Matsushita},
  {Vila-Vilar{\'o}}, \& {Kawabe}}]{2003PASJ...55L...1K}
{Kohno}, K., {Ishizuki}, S., {Matsushita}, S., {Vila-Vilar{\'o}}, B., \&
  {Kawabe}, R. 2003, \pasj, 55, L1

\bibitem[{{Kohno} {et~al.}(2008){Kohno}, {Nakanishi}, {Tosaki}, {Muraoka},
  {Miura}, {Ezawa}, \& {Kawabe}}]{2008Ap&SS.313..279K}
{Kohno}, K., {Nakanishi}, K., {Tosaki}, T., {et~al.} 2008, \apss, 313, 279

\bibitem[{{Kormendy} \& {Ho}(2013)}]{2013ARA&A..51..511K}
{Kormendy}, J., \& {Ho}, L.~C. 2013, \araa, 51, 511

\bibitem[{{Kormendy} \& {Kennicutt}(2004)}]{2004ARA&A..42..603K}
{Kormendy}, J., \& {Kennicutt}, Jr., R.~C. 2004, \araa, 42, 603

\bibitem[{{Kraemer} {et~al.}(2012){Kraemer}, {Crenshaw}, {Dunn}, {Turner},
  {Lobban}, {Miller}, {Reeves}, {Fischer}, \& {Braito}}]{2012ApJ...751...84K}
{Kraemer}, S.~B., {Crenshaw}, D.~M., {Dunn}, J.~P., {et~al.} 2012, \apj, 751,
  84

\bibitem[{{Krips} {et~al.}(2008){Krips}, {Neri}, {Garc{\'{\i}}a-Burillo},
  {Mart{\'{\i}}n}, {Combes}, {Graci{\'a}-Carpio}, \&
  {Eckart}}]{2008ApJ...677..262K}
{Krips}, M., {Neri}, R., {Garc{\'{\i}}a-Burillo}, S., {et~al.} 2008, \apj, 677,
  262

\bibitem[{{Krips} {et~al.}(2007){Krips}, {Neri}, {Garc{\'{\i}}a-Burillo},
  {Combes}, {Schinnerer}, {Baker}, {Eckart}, {Boone}, {Hunt}, {Leon}, \&
  {Tacconi}}]{2007A&A...468L..63K}
---. 2007, \aap, 468, L63

\bibitem[{{Krongold} {et~al.}(2007){Krongold}, {Nicastro}, {Elvis},
  {Brickhouse}, {Binette}, {Mathur}, \&
  {Jim{\'e}nez-Bail{\'o}n}}]{2007ApJ...659.1022K}
{Krongold}, Y., {Nicastro}, F., {Elvis}, M., {et~al.} 2007, \apj, 659, 1022

\bibitem[{{Kuo} {et~al.}(2011){Kuo}, {Braatz}, {Condon}, {Impellizzeri}, {Lo},
  {Zaw}, {Schenker}, {Henkel}, {Reid}, \& {Greene}}]{2011ApJ...727...20K}
{Kuo}, C.~Y., {Braatz}, J.~A., {Condon}, J.~J., {et~al.} 2011, \apj, 727, 20

\bibitem[{{Lepp} \& {Dalgarno}(1996)}]{1996A&A...306L..21L}
{Lepp}, S., \& {Dalgarno}, A. 1996, \aap, 306, L21

\bibitem[{{Lin} {et~al.}(2016){Lin}, {Davies}, {Burtscher}, {Contursi},
  {Genzel}, {Gonz{\'a}lez-Alfonso}, {Graci{\'a}-Carpio}, {Janssen}, {Lutz},
  {Orban de Xivry}, {Rosario}, {Schnorr-M{\"u}ller}, {Sternberg}, {Sturm}, \&
  {Tacconi}}]{2016MNRAS.458.1375L}
{Lin}, M.-Y., {Davies}, R.~I., {Burtscher}, L., {et~al.} 2016, \mnras, 458,
  1375

\bibitem[{{Liu} {et~al.}(2014){Liu}, {Wang}, {Yang}, {Zhu}, \&
  {Zhou}}]{2014ApJ...783..106L}
{Liu}, T., {Wang}, J.-X., {Yang}, H., {Zhu}, F.-F., \& {Zhou}, Y.-Y. 2014,
  \apj, 783, 106

\bibitem[{{Lodato} \& {Bertin}(2003)}]{2003A&A...398..517L}
{Lodato}, G., \& {Bertin}, G. 2003, \aap, 398, 517

\bibitem[{{Lutz} {et~al.}(2004){Lutz}, {Maiolino}, {Spoon}, \&
  {Moorwood}}]{2004A&A...418..465L}
{Lutz}, D., {Maiolino}, R., {Spoon}, H.~W.~W., \& {Moorwood}, A.~F.~M. 2004,
  \aap, 418, 465

\bibitem[{{Machida} \& {Matsumoto}(2003)}]{2003ApJ...585..429M}
{Machida}, M., \& {Matsumoto}, R. 2003, \apj, 585, 429

\bibitem[{{Magorrian} {et~al.}(1998){Magorrian}, {Tremaine}, {Richstone},
  {Bender}, {Bower}, {Dressler}, {Faber}, {Gebhardt}, {Green}, {Grillmair},
  {Kormendy}, \& {Lauer}}]{1998AJ....115.2285M}
{Magorrian}, J., {Tremaine}, S., {Richstone}, D., {et~al.} 1998, \aj, 115, 2285

\bibitem[{{Maiolino} \& {Rieke}(1995)}]{1995ApJ...454...95M}
{Maiolino}, R., \& {Rieke}, G.~H. 1995, \apj, 454, 95

\bibitem[{{Makarov} {et~al.}(2014){Makarov}, {Prugniel}, {Terekhova},
  {Courtois}, \& {Vauglin}}]{2014A&A...570A..13M}
{Makarov}, D., {Prugniel}, P., {Terekhova}, N., {Courtois}, H., \& {Vauglin},
  I. 2014, \aap, 570, A13

\bibitem[{{Maloney} {et~al.}(1996){Maloney}, {Hollenbach}, \&
  {Tielens}}]{1996ApJ...466..561M}
{Maloney}, P.~R., {Hollenbach}, D.~J., \& {Tielens}, A.~G.~G.~M. 1996, \apj,
  466, 561

\bibitem[{{Marconi} \& {Hunt}(2003)}]{2003ApJ...589L..21M}
{Marconi}, A., \& {Hunt}, L.~K. 2003, \apjl, 589, L21

\bibitem[{{Marconi} {et~al.}(2004){Marconi}, {Risaliti}, {Gilli}, {Hunt},
  {Maiolino}, \& {Salvati}}]{2004MNRAS.351..169M}
{Marconi}, A., {Risaliti}, G., {Gilli}, R., {et~al.} 2004, \mnras, 351, 169

\bibitem[{{Marinucci} {et~al.}(2012){Marinucci}, {Bianchi}, {Nicastro}, {Matt},
  \& {Goulding}}]{2012ApJ...748..130M}
{Marinucci}, A., {Bianchi}, S., {Nicastro}, F., {Matt}, G., \& {Goulding},
  A.~D. 2012, \apj, 748, 130

\bibitem[{{Mart{\'{\i}}n} {et~al.}(2015){Mart{\'{\i}}n}, {Kohno}, {Izumi},
  {Krips}, {Meier}, {Aladro}, {Matsushita}, {Takano}, {Turner}, {Espada},
  {Nakajima}, {Terashima}, {Fathi}, {Hsieh}, {Imanishi}, {Lundgren}, {Nakai},
  {Schinnerer}, {Sheth}, \& {Wiklind}}]{2015A&A...573A.116M}
{Mart{\'{\i}}n}, S., {Kohno}, K., {Izumi}, T., {et~al.} 2015, \aap, 573, A116

\bibitem[{{Martini} {et~al.}(2003){Martini}, {Regan}, {Mulchaey}, \&
  {Pogge}}]{2003ApJ...589..774M}
{Martini}, P., {Regan}, M.~W., {Mulchaey}, J.~S., \& {Pogge}, R.~W. 2003, \apj,
  589, 774

\bibitem[{{Matsushita} {et~al.}(2015){Matsushita}, {Trung}, {Boone}, {Krips},
  {Lim}, \& {Muller}}]{2015ApJ...799...26M}
{Matsushita}, S., {Trung}, D.-V., {Boone}, F., {et~al.} 2015, \apj, 799, 26

\bibitem[{{McMullin} {et~al.}(2007){McMullin}, {Waters}, {Schiebel}, {Young},
  \& {Golap}}]{2007ASPC..376..127M}
{McMullin}, J.~P., {Waters}, B., {Schiebel}, D., {Young}, W., \& {Golap}, K.
  2007, in Astronomical Society of the Pacific Conference Series, Vol. 376,
  Astronomical Data Analysis Software and Systems XVI, ed. R.~A. {Shaw},
  F.~{Hill}, \& D.~J. {Bell}, 127

\bibitem[{{Mihos} \& {Hernquist}(1994)}]{1994ApJ...425L..13M}
{Mihos}, J.~C., \& {Hernquist}, L. 1994, \apjl, 425, L13

\bibitem[{{Miyoshi} {et~al.}(1995){Miyoshi}, {Moran}, {Herrnstein},
  {Greenhill}, {Nakai}, {Diamond}, \& {Inoue}}]{1995Natur.373..127M}
{Miyoshi}, M., {Moran}, J., {Herrnstein}, J., {et~al.} 1995, \nat, 373, 127

\bibitem[{{Monje} {et~al.}(2011){Monje}, {Blain}, \&
  {Phillips}}]{2011ApJS..195...23M}
{Monje}, R.~R., {Blain}, A.~W., \& {Phillips}, T.~G. 2011, \apjs, 195, 23

\bibitem[{{Mortlock} {et~al.}(2011){Mortlock}, {Warren}, {Venemans}, {Patel},
  {Hewett}, {McMahon}, {Simpson}, {Theuns}, {Gonz{\'a}les-Solares}, {Adamson},
  {Dye}, {Hambly}, {Hirst}, {Irwin}, {Kuiper}, {Lawrence}, \&
  {R{\"o}ttgering}}]{2011Natur.474..616M}
{Mortlock}, D.~J., {Warren}, S.~J., {Venemans}, B.~P., {et~al.} 2011, \nat,
  474, 616

\bibitem[{{Mulchaey} \& {Regan}(1997)}]{1997ApJ...482L.135M}
{Mulchaey}, J.~S., \& {Regan}, M.~W. 1997, \apjl, 482, L135

\bibitem[{{M{\"u}ller S{\'a}nchez} {et~al.}(2009){M{\"u}ller S{\'a}nchez},
  {Davies}, {Genzel}, {Tacconi}, {Eisenhauer}, {Hicks}, {Friedrich}, \&
  {Sternberg}}]{2009ApJ...691..749M}
{M{\"u}ller S{\'a}nchez}, F., {Davies}, R.~I., {Genzel}, R., {et~al.} 2009,
  \apj, 691, 749

\bibitem[{{Narayan} \& {Yi}(1995)}]{1995ApJ...452..710N}
{Narayan}, R., \& {Yi}, I. 1995, \apj, 452, 710

\bibitem[{{Neumayer}(2010)}]{2010PASA...27..449N}
{Neumayer}, N. 2010, \pasa, 27, 449

\bibitem[{{Norman} \& {Scoville}(1988)}]{1988ApJ...332..124N}
{Norman}, C., \& {Scoville}, N. 1988, \apj, 332, 124

\bibitem[{{Onishi} {et~al.}(2015){Onishi}, {Iguchi}, {Sheth}, \&
  {Kohno}}]{2015ApJ...806...39O}
{Onishi}, K., {Iguchi}, S., {Sheth}, K., \& {Kohno}, K. 2015, \apj, 806, 39

\bibitem[{{Padovani} \& {Matteucci}(1993)}]{1993ApJ...416...26P}
{Padovani}, P., \& {Matteucci}, F. 1993, \apj, 416, 26

\bibitem[{{Panessa} {et~al.}(2006){Panessa}, {Bassani}, {Cappi}, {Dadina},
  {Barcons}, {Carrera}, {Ho}, \& {Iwasawa}}]{2006A&A...455..173P}
{Panessa}, F., {Bassani}, L., {Cappi}, M., {et~al.} 2006, \aap, 455, 173

\bibitem[{{Papadopoulos} {et~al.}(2012){Papadopoulos}, {van der Werf},
  {Xilouris}, {Isaak}, \& {Gao}}]{2012ApJ...751...10P}
{Papadopoulos}, P.~P., {van der Werf}, P., {Xilouris}, E., {Isaak}, K.~G., \&
  {Gao}, Y. 2012, \apj, 751, 10

\bibitem[{{Petry} \& {CASA Development Team}(2012)}]{2012ASPC..461..849P}
{Petry}, D., \& {CASA Development Team}. 2012, in Astronomical Society of the
  Pacific Conference Series, Vol. 461, Astronomical Data Analysis Software and
  Systems XXI, ed. P.~{Ballester}, D.~{Egret}, \& N.~P.~F. {Lorente}, 849

\bibitem[{{Pringle}(1981)}]{1981ARA&A..19..137P}
{Pringle}, J.~E. 1981, \araa, 19, 137

\bibitem[{{Privon} {et~al.}(2015){Privon}, {Herrero-Illana}, {Evans},
  {Iwasawa}, {Perez-Torres}, {Armus}, {D{\'{\i}}az-Santos}, {Murphy},
  {Stierwalt}, {Aalto}, {Mazzarella}, {Barcos-Mu{\~n}oz}, {Borish}, {Inami},
  {Kim}, {Treister}, {Surace}, {Lord}, {Conway}, {Frayer}, \&
  {Alberdi}}]{2015ApJ...814...39P}
{Privon}, G.~C., {Herrero-Illana}, R., {Evans}, A.~S., {et~al.} 2015, \apj,
  814, 39

\bibitem[{{Rigby} {et~al.}(2009){Rigby}, {Diamond-Stanic}, \&
  {Aniano}}]{2009ApJ...700.1878R}
{Rigby}, J.~R., {Diamond-Stanic}, A.~M., \& {Aniano}, G. 2009, \apj, 700, 1878

\bibitem[{{Rocca-Volmerange} {et~al.}(2015){Rocca-Volmerange}, {Drouart}, \&
  {De Breuck}}]{2015ApJ...803L...8R}
{Rocca-Volmerange}, B., {Drouart}, G., \& {De Breuck}, C. 2015, \apjl, 803, L8

\bibitem[{{Sakamoto} {et~al.}(1999){Sakamoto}, {Okumura}, {Ishizuki}, \&
  {Scoville}}]{1999ApJ...525..691S}
{Sakamoto}, K., {Okumura}, S.~K., {Ishizuki}, S., \& {Scoville}, N.~Z. 1999,
  \apj, 525, 691

\bibitem[{{Sandage} \& {Tammann}(1987)}]{1987rsac.book.....S}
{Sandage}, A., \& {Tammann}, G.~A. 1987, {A revised Shapley-Ames Catalog of
  Bright Galaxies} (2nd ed.; Washington, DC: Carnegie Institution of Washington
  Publication)

\bibitem[{{Sani} {et~al.}(2012){Sani}, {Davies}, {Sternberg},
  {Graci{\'a}-Carpio}, {Hicks}, {Krips}, {Tacconi}, {Genzel}, {Vollmer},
  {Schinnerer}, {Garc{\'{\i}}a-Burillo}, {Usero}, \& {Orban de
  Xivry}}]{2012MNRAS.424.1963S}
{Sani}, E., {Davies}, R.~I., {Sternberg}, A., {et~al.} 2012, \mnras, 424, 1963

\bibitem[{{Sault} {et~al.}(1995){Sault}, {Teuben}, \&
  {Wright}}]{1995ASPC...77..433S}
{Sault}, R.~J., {Teuben}, P.~J., \& {Wright}, M.~C.~H. 1995, in Astronomical
  Society of the Pacific Conference Series, Vol.~77, Astronomical Data Analysis
  Software and Systems IV, ed. R.~A. {Shaw}, H.~E. {Payne}, \& J.~J.~E.
  {Hayes}, 433

\bibitem[{{Schawinski} {et~al.}(2009){Schawinski}, {Virani}, {Simmons}, {Urry},
  {Treister}, {Kaviraj}, \& {Kushkuley}}]{2009ApJ...692L..19S}
{Schawinski}, K., {Virani}, S., {Simmons}, B., {et~al.} 2009, \apjl, 692, L19

\bibitem[{{Shakura} \& {Sunyaev}(1973)}]{1973A&A....24..337S}
{Shakura}, N.~I., \& {Sunyaev}, R.~A. 1973, \aap, 24, 337

\bibitem[{{Sheth} {et~al.}(2005){Sheth}, {Vogel}, {Regan}, {Thornley}, \&
  {Teuben}}]{2005ApJ...632..217S}
{Sheth}, K., {Vogel}, S.~N., {Regan}, M.~W., {Thornley}, M.~D., \& {Teuben},
  P.~J. 2005, \apj, 632, 217

\bibitem[{{Shlosman} {et~al.}(1990){Shlosman}, {Begelman}, \&
  {Frank}}]{1990Natur.345..679S}
{Shlosman}, I., {Begelman}, M.~C., \& {Frank}, J. 1990, \nat, 345, 679

\bibitem[{{Sim{\~o}es Lopes} {et~al.}(2007){Sim{\~o}es Lopes},
  {Storchi-Bergmann}, {de F{\'a}tima Saraiva}, \&
  {Martini}}]{2007ApJ...655..718S}
{Sim{\~o}es Lopes}, R.~D., {Storchi-Bergmann}, T., {de F{\'a}tima Saraiva}, M.,
  \& {Martini}, P. 2007, \apj, 655, 718

\bibitem[{{Solomon} {et~al.}(1992{\natexlab{a}}){Solomon}, {Downes}, \&
  {Radford}}]{1992ApJ...387L..55S}
{Solomon}, P.~M., {Downes}, D., \& {Radford}, S.~J.~E. 1992{\natexlab{a}},
  \apjl, 387, L55

\bibitem[{{Solomon} {et~al.}(1992{\natexlab{b}}){Solomon}, {Downes}, \&
  {Radford}}]{1992ApJ...398L..29S}
---. 1992{\natexlab{b}}, \apjl, 398, L29

\bibitem[{{Solomon} {et~al.}(1990){Solomon}, {Radford}, \&
  {Downes}}]{1990ApJ...348L..53S}
{Solomon}, P.~M., {Radford}, S.~J.~E., \& {Downes}, D. 1990, \apjl, 348, L53

\bibitem[{{Solomon} \& {Vanden Bout}(2005)}]{2005ARA&A..43..677S}
{Solomon}, P.~M., \& {Vanden Bout}, P.~A. 2005, \araa, 43, 677

\bibitem[{{Taniguchi}(1999)}]{1999ApJ...524...65T}
{Taniguchi}, Y. 1999, \apj, 524, 65

\bibitem[{{Thompson} {et~al.}(2005){Thompson}, {Quataert}, \&
  {Murray}}]{2005ApJ...630..167T}
{Thompson}, T.~A., {Quataert}, E., \& {Murray}, N. 2005, \apj, 630, 167

\bibitem[{{Toomre}(1964)}]{1964ApJ...139.1217T}
{Toomre}, A. 1964, \apj, 139, 1217

\bibitem[{{Treister} {et~al.}(2012){Treister}, {Schawinski}, {Urry}, \&
  {Simmons}}]{2012ApJ...758L..39T}
{Treister}, E., {Schawinski}, K., {Urry}, C.~M., \& {Simmons}, B.~D. 2012,
  \apjl, 758, L39

\bibitem[{{Tremaine} {et~al.}(2002){Tremaine}, {Gebhardt}, {Bender}, {Bower},
  {Dressler}, {Faber}, {Filippenko}, {Green}, {Grillmair}, {Ho}, {Kormendy},
  {Lauer}, {Magorrian}, {Pinkney}, \& {Richstone}}]{2002ApJ...574..740T}
{Tremaine}, S., {Gebhardt}, K., {Bender}, R., {et~al.} 2002, \apj, 574, 740

\bibitem[{{Ueda} {et~al.}(2014){Ueda}, {Akiyama}, {Hasinger}, {Miyaji}, \&
  {Watson}}]{2014ApJ...786..104U}
{Ueda}, Y., {Akiyama}, M., {Hasinger}, G., {Miyaji}, T., \& {Watson}, M.~G.
  2014, \apj, 786, 104

\bibitem[{{Ueda} {et~al.}(2003){Ueda}, {Akiyama}, {Ohta}, \&
  {Miyaji}}]{2003ApJ...598..886U}
{Ueda}, Y., {Akiyama}, M., {Ohta}, K., \& {Miyaji}, T. 2003, \apj, 598, 886

\bibitem[{{Urry} \& {Padovani}(1995)}]{1995PASP..107..803U}
{Urry}, C.~M., \& {Padovani}, P. 1995, \pasp, 107, 803

\bibitem[{{Viti} {et~al.}(2014){Viti}, {Garc{\'{\i}}a-Burillo}, {Fuente},
  {Hunt}, {Usero}, {Henkel}, {Eckart}, {Martin}, {Spaans}, {Muller}, {Combes},
  {Krips}, {Schinnerer}, {Casasola}, {Costagliola}, {Marquez}, {Planesas}, {van
  der Werf}, {Aalto}, {Baker}, {Boone}, \& {Tacconi}}]{2014A&A...570A..28V}
{Viti}, S., {Garc{\'{\i}}a-Burillo}, S., {Fuente}, A., {et~al.} 2014, \aap,
  570, A28

\bibitem[{{Vollmer} {et~al.}(2008){Vollmer}, {Beckert}, \&
  {Davies}}]{2008A&A...491..441V}
{Vollmer}, B., {Beckert}, T., \& {Davies}, R.~I. 2008, \aap, 491, 441

\bibitem[{{Wada} \& {Norman}(2002)}]{2002ApJ...566L..21W}
{Wada}, K., \& {Norman}, C.~A. 2002, \apjl, 566, L21

\bibitem[{{Wada} {et~al.}(2009){Wada}, {Papadopoulos}, \&
  {Spaans}}]{2009ApJ...702...63W}
{Wada}, K., {Papadopoulos}, P.~P., \& {Spaans}, M. 2009, \apj, 702, 63

\bibitem[{{Wang} {et~al.}(2004){Wang}, {Henkel}, {Chin}, {Whiteoak}, {Hunt
  Cunningham}, {Mauersberger}, \& {Muders}}]{2004A&A...422..883W}
{Wang}, M., {Henkel}, C., {Chin}, Y.-N., {et~al.} 2004, \aap, 422, 883

\bibitem[{{Wang} {et~al.}(2011){Wang}, {Wagg}, {Carilli}, {Walter}, {Riechers},
  {Willott}, {Bertoldi}, {Omont}, {Beelen}, {Cox}, {Strauss}, {Bergeron},
  {Forveille}, {Menten}, \& {Fan}}]{2011ApJ...739L..34W}
{Wang}, R., {Wagg}, J., {Carilli}, C.~L., {et~al.} 2011, \apjl, 739, L34

\bibitem[{{Wild} {et~al.}(2010){Wild}, {Heckman}, \&
  {Charlot}}]{2010MNRAS.405..933W}
{Wild}, V., {Heckman}, T., \& {Charlot}, S. 2010, \mnras, 405, 933

\bibitem[{{Wild} \& {Eckart}(2000)}]{2000A&A...359..483W}
{Wild}, W., \& {Eckart}, A. 2000, \aap, 359, 483

\bibitem[{{Wu} {et~al.}(2005){Wu}, {Evans}, {Gao}, {Solomon}, {Shirley}, \&
  {Vanden Bout}}]{2005ApJ...635L.173W}
{Wu}, J., {Evans}, II, N.~J., {Gao}, Y., {et~al.} 2005, \apjl, 635, L173

\bibitem[{{Yamada}(1994)}]{1994ApJ...423L..27Y}
{Yamada}, T. 1994, \apjl, 423, L27

\end{thebibliography}

\appendix
\section{A. Archival data of NGC 4579}\label{app-A}
NGC 4579 (luminosity distance = 21.7 Mpc) was observed with ALMA on 2014 April 25 
with 37 antennas as a Cycle 1 early science program 
(ID = 2012.1.00456.S, PI = E. Murphy). 
Baseline lengths range from 16.4 m to 558.2 m. 
The observations were conducted in a single pointing with a $\sim$60$''$ field of view, 
which fully covered the nuclear region of this galaxy. 
The phase tracking center was set to 
($\alpha_{\rm J2000.0}$, $\delta_{\rm J2000.0}$) 
= (12$^{\rm h}$37$^{\rm m}$43.6$^{\rm s}$, +11$^\circ$49$'$02.0$''$). 
The HCN($1-0)$ emission line was covered in the lower side band (LSB), 
whereas the upper side band (USB) was used to improve the sensitivity 
to the underlying continuum emission. 
Both side bands in total covered $\sim$7.5 GHz width. 
The velocity spacing was originally $\sim$1.65 km s$^{-1}$ (488.281 kHz), 
but we binned 30 channels to achieve 50 km s$^{-1}$ resolution to improve the signal to noise ratio. 
The bandpass, phase, and absolute flux were calibrated with 
J1229+0203, J1239+0730, and Ceres, respectively. 
The total on-source time was $\sim$0.2 hr 
and the typical system temperature was $\sim$50$-75$ K. 

\begin{figure}
\epsscale{1}
\plotone{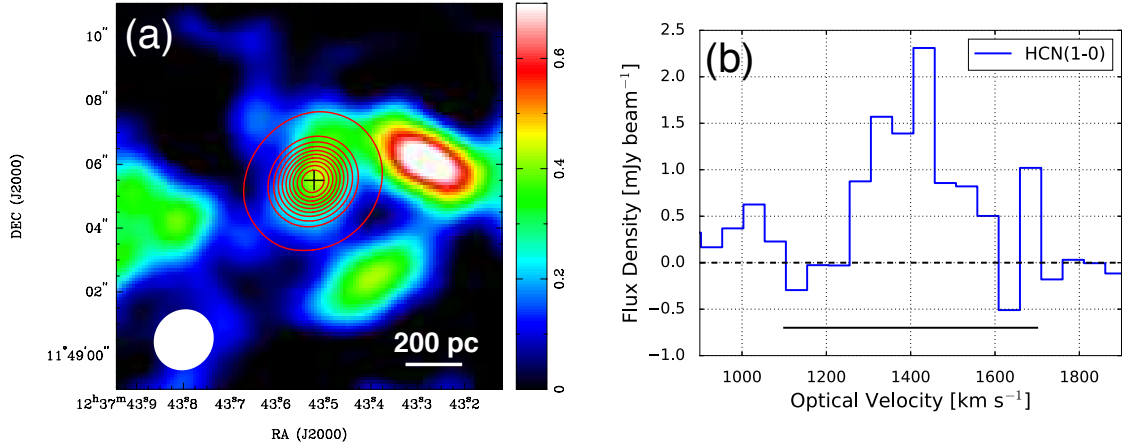}
\caption{
(a) Spatial distributions of the HCN($1-0$) emission (color) and the underlying 95 GHz continuum emission (contour) of the central 1.2 kpc region of NGC 4579. 
The central cross indicates the AGN position. 
The contours are, 10, 100, 150, 200, 250, 300, 350, 400, 450, and 500$\sigma$, where 1$\sigma$ = 0.196 mJy beam$^{-1}$ (maximum = 109 mJy beam$^{-1}$). 
The 1$\sigma$ noise level in the HCN($1-0$) integrated intensity map is 0.126 Jy beam$^{-1}$ km s$^{-1}$. 
The value at the AGN position is 0.437 Jy beam$^{-1}$ km s$^{-1}$ ($\sim$3.5$\sigma$). 
The bottom-left white ellipse indicates the synthesized beam for the HCN($1-0$) emission (1$''$.96 $\times$ 1$''$.83, P.A. = $-$37.75$^\circ$). 
(b) The HCN($1-0$) spectrum extracted at the AGN position with a single synthesized beam. 
We integrated over the velocity range of 1100$-$1700 km s$^{-1}$ 
as indicated by the horizontal black, solid line, 
to make the moment-0 map shown in (a). 
}
\label{figureA1}
\end{figure}

The reduction and calibration of the data were conducted with CASA version 4.2 (\citealt{2007ASPC..376..127M,2012ASPC..461..849P}) in standard manners. 
The raw visibility data was extracted from the ALMA archive\footnote{\url{https://almascience.nao.ac.jp/alma-data/archive}}. 
The HCN($1-0$) data cube was reconstructed with the CASA task \verb|CLEAN| 
(gain = 0.1, threshold = 0.5 mJy, weighting = natural). 
The achieved synthesized beam was 1$''$.96 $\times$ 1$''$.83 (207 pc $\times$ 193 pc) with P.A. = $-$37.75$^\circ$. 
The rms noise in the channel map was $\sim$0.70 mJy beam$^{-1}$ after the primary beam correction, 
which was measured at the areas free of line emission but close to the line emitting regions. 
The underlying continuum emission ($\nu_{\rm rest} = 95$ GHz; synthesized beam was 1$''$.89 $\times$ 1$''$.68, P.A. = $-$43.39$^\circ$) 
was subtracted in the $uv$ plane before making the line map. 
The rms noise in this continuum map was 0.20 mJy beam$^{-1}$. 
We further analyzed this line cube with \verb|MIRIAD| (\citealt{1995ASPC...77..433S}). 
Figure \ref{figureA1} shows the integrated intensity map of HCN($1-0$) 
(we used the \verb|MIRIAD| task \verb|MOMENT| to make this map without any clipping) and its spectrum extracted at the nucleus. 
The moment-0 value at the nucleus (0.437$\pm$0.126 Jy beam$^{-1}$ km s$^{-1}$) was used to derive $L'_{\rm HCN}$ and $M_{\rm dense}$ in this work. 

\end{document}